\documentclass[pre,twocolumn,dcolumn,superscriptaddress,showpacs]{revtex4}

\usepackage{hyperref}
\usepackage{psfrag}

\usepackage{bm,amsmath}
\usepackage{graphicx}
\usepackage{color}

\renewcommand{\d}{{d}}

\newcommand{\gdot}{\dot{\gamma}}

\begin{document}

\title{The glass susceptibility: growth kinetics and saturation under shear}

\author{Saroj Kumar Nandi}
\email{saroj.nandi@cea.fr}
\affiliation{Centre for Condensed Matter Theory, Department of Physics, Indian Institute of Science, Bangalore - 560 012, India}
\affiliation{IPhT, CEA/DSM-CNRS/URA 2306, CEA Saclay, F-91191 Gif-sur-Yvette Cedex, France\footnote{Present address}}

\author{Sriram Ramaswamy}
\email{sriram@tifrh.res.in}
\affiliation{Centre for Condensed Matter Theory, Department of Physics, Indian Institute of Science, Bangalore - 560 012, India}
\affiliation{TIFR Centre for Interdisciplinary Sciences, 21 Brundavan Colony, Narsingi, Hyderabad 500 075, India}

\begin{abstract}
We study the growth kinetics of glassy correlations in a structural glass by
monitoring the evolution, within mode-coupling theory, of a suitably defined
three-point function $\chi_C(t,t_w)$ with time $t$ and waiting time $t_w$. From
the complete wave vector-dependent equations of motion for domain growth we pass
to a schematic limit to obtain a numerically tractable form.
We find that the peak value $\chi_C^P$ of $\chi_C(t,t_w)$, which can be viewed
as a correlation volume, grows as $t_w^{0.5}$, and
the relaxation time as $t_w^{0.8}$, following a quench to a point deep in the
glassy state. These results constitute a theoretical explanation of the
simulation findings of Parisi [J. Phys. Chem. B {\bf 103}, 4128 (1999)] and Kob
and Barrat [Phys. Rev. Lett. {\bf 78}, 4581 (1997)] and are also in qualitative
agreement with Parsaeian and Castillo [Phys. Rev. E {\bf 78}, 060105(R) (2008)].
On the other hand, if the quench is to a point on the {\em liquid side}, the
correlation volume grows to saturation. We present a similar calculation for the
growth kinetics in a $p$-spin spin glass mean-field model where we find a slower
growth, $\chi_C^P \sim t_w^{0.13}$.
Further, we show that a shear rate $\gdot$ cuts off the growth of glassy
correlations when $t_w\sim 1/\gdot$ for quench in the glassy regime and
$t_w=\min(t_r,1/\gdot)$ in the liquid, where $t_r$ is the relaxation time of the
unsheared liquid. The relaxation time of the steady state fluid in this
case is $\propto \gdot^{-0.8}$.
\end{abstract}


\maketitle

\section{Introduction}

\subsection{Background}

In systems in which the formation of the equilibrium crystalline phase is easily evaded, and a glass forms without rapid cooling, the liquid-glass transition can usefully be viewed as a thermodynamic transition. The order parameter that distinguishes a glass from a liquid, as established many years ago \cite{EA75,mezard88,kobbook} by analogy with the case of a spin glass, is the time-persistent part of the density auto-correlation function. The corresponding susceptibility which measures correlations of glassiness must involve four densities \cite{chandan91}. An intense search using such higher order correlators has established in theories \cite{franz00,corberi10,biroli06,berthier07a,berthier07b}, ``equilibrium'' experiments \cite{fragiadakis11,hong11} and simulations \cite{rotman10,karmakar09}, the existence of a dynamic length scale that grows upon approaching the glass transition. In conventional critical phenomena a single diverging correlation length governs the critical-point singularities in various quantities such as order parameter, susceptibility and specific heat \cite{chaikin_lubenski,hohenberg_halperin}. For glass, several length scales have been defined \cite{lubchenko07,adam65,gibbs58,kurchan11,chiara12,karmakar09,chandan91,biroli06,biroli04,karmakar_procaccia,bouchaudbiroli04} whose inter-relations or independence are a subject of active discussion \cite{chandan_review}. Our analysis in this paper concerns the extension of the dynamic length scale, extracted from three- or four-density correlators, to the non-stationary regime following quench. By {\em equilibrium} in this paper, we mean without a quench. We shall assume we are working with good glass formers that display a glass transition independent of cooling rate.

If we treat glass as a phase, with an order parameter, we can then ask how glassiness grows following a quench. The theory of the domain growth of an ordered phase after sudden quench from the disordered phase is one of the landmark achievements of nonequilibrium statistical mechanics \cite{bray94}. 
Analogously, if there exists a length scale describing the spatial extent of glassy correlations, it too must grow as one waits longer in the final quenched state. The issue for glass was first explored by Parisi in the context of a Monte Carlo simulation of a binary mixture of soft spheres \cite{parisi99}. Such a growth of a length scale was also found, by Parsaeian {\it et al} \cite{parsaeian08} in their study of the domain growth dynamics of glassy order within the molecular dynamics simulation of a binary Lennard-Jones system. However, a detailed theoretical understanding of these findings, that is, a theory of the growth kinetics of a glass, has emerged only recently \cite{saroj12}, in an MCT framework.

Aging in structural glasses has been investigated through the study of the two-point correlator in experiments \cite{bonn02,bonn04,di11}, simulations \cite{kob97,kob00,parsaeian08}, within mode-coupling theories \cite{reichman05,castellani05,bouchaud96,latz00,latz02,gregorio02} and within Random First Order Transition (RFOT) theory \cite{lubchenko04,wolynes09}. Related studies for spin glasses include \cite{cugliandolo93,cugliandolo94,kim01,sunil09,alvarez10,chiara12}. Franz and Hertz \cite{franz95} showed that the out-of-equilibrium dynamics of the Amit-Roginsky $\phi^3$ model \cite{amit79} contains the aging dynamics observed in structural glasses and in many spin glasses. 

Mode-coupling theory (MCT) has been remarkably successful in describing glassy dynamics, notwithstanding the fact that the ``MCT glass transition'' to a non-ergodic state is ultimately avoided in real systems as a result of activated processes. Taking the input of the static structure factor alone, MCT offers parameter-free predictions of the dynamics and growth of relaxation time of dense liquids at equilibrium. Therefore, it becomes imperative to extend MCT to the case of an aging system as the first step towards a theory of the coarsening of glassiness. However, obtaining the equations of motion for the aging regime poses challenges, as time translation invariance is lost. Most conventional approaches \cite{zaccarelli02,reichman05,kawasaki03,spdas04,miyazaki05} to derive MCT use the fluctuation-dissipation relation (FDR) at some point, explicitly or implicitly. The field-theoretical technique \cite{castellani05,reichman05,bouchaud96,mythesis,barabasi_stanley} is especially well suited for this purpose as it does not assume the FDR. We use this technique to obtain the final equations for correlation and response functions starting with the hydrodynamic equations of motion. The problem of satisfying the equilibrium FDR within this approach at one-loop order has been extensively discussed \cite{andreanov06,abhik07,kim07,kim08}. But we are interested in the schematic version of the theory, within which there is no problem. Moreover, if we impose FDR by hand on the final equations, we find they reproduce the equilibrium results.

As the theoretical system size is infinite, the dynamics following a quench in our theory will be characterized by a correlation length and total susceptibility that will grow forever, as is familiar from the domain growth of a conventional ordered phase \cite{bray94}. Within our calculation, of course, the phase in question is the ``MCT glass''. However, if we apply a small shear on the system, it will reach a steady state as the waiting time becomes of the order of the inverse shear rate. Thus the dynamic length scale in glassy system under shear is restricted by the imposed shear rate.
Within MCT, shear has two primary effects on the system: (i) it reduces the height of the static structure factor, which becomes anisotropic under shear \cite{indrani95,kuni02,kuni04} and (ii) due to advection of wave vector, the strength of the memory kernel diminishes when the time scale becomes of the order of the waiting time \cite{fuchs03,brader07,kuni04,mythesis}. In principle, both the contributions should be taken into account. However, the first contribution makes a numerical solution of the final equations exceedingly difficult, since anisotropy increases the number of variables to be evaluated and the solution becomes hugely time consuming. We render the problem tractable by making an isotropic approximation \cite{fuchs03,kuni04,mizuno11} within which only the reduction in the memory kernel enters.

The natural quantity to look at in order to obtain the information about a length scale is a certain four-point correlation function \cite{chandan91} because, as we remarked, the order-parameter is a two-point quantity; however, it has been demonstrated for the equilibrium case \cite{biroli06} that certain three-point correlation functions contain similar information \cite{biroli06}, and are tractable to evaluate. In practice, as was done in \cite{biroli06}, we obtain the desired quantity through a suitably defined susceptibility.

\subsection{Results}
\begin{figure}
\begin{center}
\includegraphics[width=8.6cm]{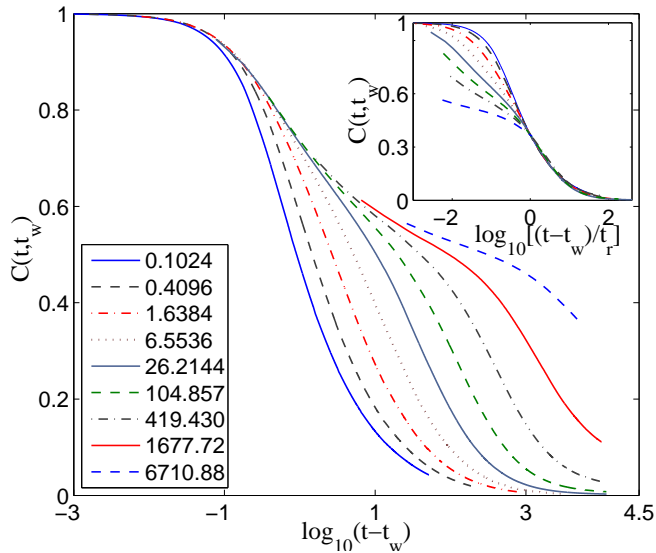}
\end{center}
\caption{The correlation function $C(t,t_w)$ is shown as a function of $t-t_w$, for various waiting times $t_w$ shown in the legend. The decay with $(t-t_w)$ becomes progressively slower with
increasing $t_w$. The final parameter values are $T=1.0$ and $\lambda=2.01$.
{\bf Inset:} Scaling $t-t_w$ by $t_r$ (see text for the definition) yields a data collapse in the $\alpha$-relaxation regime; this is the characteristic of ``simple aging''. The waiting times for various curves are same as in the main figure. However, no such data collapse is seen in the behaviour of the three-point function, Fig. \ref{scaling_fit}.}
\label{correlationfn}
\end{figure}

\begin{figure}
\begin{center}
\includegraphics[width=7.6cm]{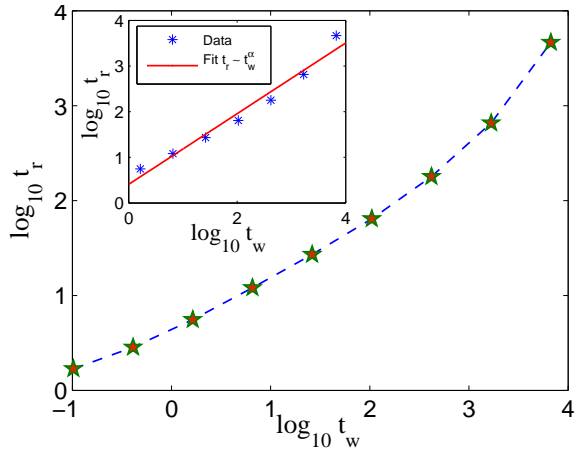}
\end{center}
\caption{The relaxation time $t_r$, defined as the time when the correlation function becomes $1/e$, as a function of waiting time. {\bf Inset:} $t_r$ for large $t_w$ can be fitted with an algebraic form $t_r\sim t_w^\alpha$ with $\alpha\approx 0.8$.}
\label{relaxationtime}
\end{figure}

The main results of this work are as follows: 
\begin{enumerate}
\item
If the quench is from the liquid state to deep in glassy regime, the peak value $\chi_C^P(t_w)$ of $\chi_C(t,t_w)$, which has the interpretation of a correlation volume, grows without bound as we wait longer in the final state (Fig. \ref{largelambda}) whereas this growth saturates when the quench is to liquid side (Fig. \ref{lowlambda}). 
\item
The correlation volume, $\chi_C^P(t_w)$ grows as $t_w^{0.5}$ and the relaxation time $t_{peak}$, defined as the time when $\chi_C(t,t_w)$ attains its peak, goes as $t_w^{0.8}$ when the quench is to glassy regime (Fig. \ref{scaling_fit} and Fig. \ref{relaxationtime}). These results rationalize the numerical experiments on domain growth \cite{parisi99} and aging \cite{kob97}. 
\item
If the quench is to a temperature still on the liquid side, the growth saturates for $t_w$ beyond the equilibrium relaxation time $t_{peak}$. The resulting finite value $\chi_C^P$ of the correlation volume goes as $\epsilon^{-1}$ where $\epsilon$ is the distance from the critical point, and as $t_{peak}^{0.56}$ (Fig. \ref{ss_chi3scaling}) when expressed in terms of the relaxation time. These results are in agreement with existing theories \cite{biroli06} and simulations \cite{karmakar09} in their appropriate limits.
\item
From the two-point function $C(t,t_w)$ we can extract a relaxation time $t_r$ where $C(t,t_w)$ becomes $1/e$. If we scale time by $t_r$, $C(t,t_w)$ shows data collapse as is expected for ``simple aging'' (Inset of Fig. \ref{correlationfn}). However, no such data collapse is seen when $\chi_C(t,t_w)$ is scaled with $\chi_C^P$ and time with $t_{peak}$ (Fig. \ref{scaling_fit}). This suggests that describing an aging system in terms of an evolving effective temperature misses some essential physics. 
\item
The mean-field model of the $p$-spin spherical spin glass is amenable to a similar treatment, and displays a much slower growth of the correlation volume, $\chi_C^P  \sim t_w^{0.13}$. 
\item
Imposing a non-zero shear rate $\gdot$ cuts off the growth of correlation volume when $t_w\sim 1/\gdot$ for quench to glassy regime and $t_w=\min(t_r,1/\gdot)$ for quench in the liquid side (Fig. \ref{ch5_gdot_2},\ref{gdot_3}). 
The relaxation time of the steady state fluid goes as $\gdot^{-0.8}$.
\end{enumerate}

A short account, presenting some of these results, appeared in \cite{saroj12}. The rest of the paper is organised as follows:  In Sec. \ref{hydrodynamic} we show the calculation for the two- and three-point correlation functions for an aging system through the field theoretic method starting from the hydrodynamic equations of motion. In Sec. \ref{schematic} we show how to obtain the aging equations for the two-point correlator and the corresponding susceptibilities within a completely schematic treatment that can also be viewed as the MCT equations for a toy Hamiltonian. We present the resulting detailed predictions of the theory in Sec. \ref{coarseningresults}. In Sec. \ref{spinglass} we outline the calculation and the corresponding results for the three-point correlator for the mean-field $p$-spin spherical spin glass model. Next, in Sec. \ref{aging_shear}, we incorporate shear into the theory of coarsening of structural glasses to see its effect on an aging system and how shear cuts off the growth of the glassy length scale. Finally we conclude the paper by discussing achievements and prospects in Sec. \ref{coarsening_discussion}.

\section{The equations of motion for an aging system}
\label{hydrodynamic}
To obtain the equations of motion governing the growth kinetics of a glassy system upon quench past the transition point, we first need to extend mode-coupling theory for the description of the two-point correlator of an aging system. We accomplish this using the field theoretic method through the hydrodynamic approach \cite{spdas04}. Let us start with the equations of hydrodynamics for a fluid with velocity field $\mathbf{v}(\mathbf{r},t)$ and density field $\rho(\mathbf{r},t)=\rho_0+\delta\rho(\mathbf{r},t)$ where $\rho_0$ is the uniform average density. The continuity equation for the density field is given by
\begin{equation}
\label{continuity}
\partial_t \rho+\nabla\cdot(\rho\mathbf{v})=0,
\end{equation}
and the generalised Navier-Stokes equation is
\begin{equation}
\label{NS}
\rho(\partial_t+\mathbf{v}\cdot\nabla)\mathbf{v}=\eta\nabla^2\mathbf{v}+(\zeta+\eta/3)\nabla\nabla\cdot\mathbf{v}-\rho\nabla\frac{\delta \mathcal{F}}{\delta\rho}+\mathbf{f},
\end{equation}
where $\eta$ and $\zeta$ are the shear and bulk viscosities, $\mathcal{F}$ is a suitably chosen density-wave free-energy functional and the thermal fluctuation is taken into the theory through the Gaussian white noise with the statistics
\begin{equation}
\label{noisestatistics}
\langle \mathbf{f}(\mathbf{0},0)\mathbf{f}(\mathbf{r},t)\rangle=-2k_BT[\eta\mathbf{I}\nabla^2+(\zeta+\eta/3)\nabla\nabla]\delta(\mathbf{r})\delta(t),
\end{equation}
where $\mathbf{I}$ is the unit tensor, $k_B$ the Boltzmann constant and $T$ the temperature. It has been shown in the literature \cite{chandan98} that the Ramakrishnan-Yussouff (RY) free energy functional \cite{ramakrishnan79} 
\begin{equation}
\label{RY}
\beta \mathcal{F}=\int d{\bf r} \left(\rho\ln\frac{\rho}{\rho_0}-\delta\rho\right)-\frac{1}{2}\int d{\bf r}d{\bf r}'c(\mathbf{r}-\mathbf{r}')\delta\rho({\bf r})\delta\rho({\bf r}'),
\end{equation}
gives a good description of ordered as well as amorphous local minima, and the corresponding dynamics in simple liquids. In Eq. (\ref{RY}) $\beta=1/k_BT$ and $c(\mathbf{r})$ is the direct pair correlation function that encodes the information of the intermolecular interactions in a coarse-grained fashion. 

We linearize the eqs. (\ref{continuity}) and (\ref{NS}), take the divergence of (\ref{NS}) and replace the divergence of the velocity field by using Eq. (\ref{continuity}). The resulting equation, after neglecting the convective nonlinearity as appropriate for a highly viscous system, will read in Fourier space as
\begin{equation}
\label{langevineq1}
\frac{\partial^2 \delta\rho_k(t)}{\partial t^2}+D_Lk^2\frac{\partial \rho_k(t)}{\partial t}=\left[ \nabla\cdot\left(\rho\nabla\frac{\delta \mathcal{F}}{\delta\rho}\right)\right]_k -i\mathbf{k}\cdot \mathbf{f}_k^L(t),
\end{equation}
where $D_L=(\zeta+4\eta/3)/\rho_0$, $\mathbf{f}_k^L(t)$ is the longitudinal part of the noise and $[\ldots]_k$ means that the term is evaluated at wave vector $k$. Ignoring the acceleration term, as we are interested in the glassy regime, and using the explicit form of the free-energy functional from Eq. (\ref{RY}), we find that the density fluctuation $\delta\rho_k(t)$ obeys
\begin{equation}
\label{ch5_eq1}
\frac{\partial \delta \rho_k(t)}{\partial t} +K_1\delta\rho_k(t)=\frac{K_2}{2}\int_{\bf q}\mathcal{V}_{k,q}\delta\rho_q(t)\delta\rho_{k-q}(t)+\hat{f}_k(t),
\end{equation}
with $\mathcal{V}_{k,q}=\mathbf{k}\cdot[\mathbf{q}c_q+(\mathbf{k}-\mathbf{q})c_{k-q}]$, $K_1={k_BT}/{S_kD_L}$ and $K_2={k_BT}/{D_Lk^2}$, $S_k$ and $c_k$ are the equilibrium structure factor and the direct correlation function respectively and the modified noise $\hat{f}_k(t)$ obeys 
\begin{equation}
\label{kdep2}
\langle \hat{f}_k(t)\hat{f}_{k'}(t')\rangle=\frac{2k_BT}{D_L}\rho_k(t)(2\pi)^d\delta(\mathbf{k}+\mathbf{k}')\delta(t-t').
\end{equation}

Eq. (\ref{ch5_eq1}) is our starting equation. We will use the diagrammatic perturbation theory technique to obtain the equations of motion for the correlation function, $C_k(t,t_w)=\langle \delta\rho_k(t)\delta\rho_{-k}(t_w)\rangle$, and response function $R_k(t,t_w)=\langle \partial \delta\rho_k(t)/\partial \eta_{-k}(t_w)\rangle$ through the field-theoretic derivation of the mode-coupling theory starting from Eq. (\ref{ch5_eq1}). The derivation is quite standard \cite{castellani05,reichman05,bouchaud96} and as we stated earlier, we skip the details. After a straightforward but tedious calculation, it is possible to write down the equations of motion for the correlation and response functions as
\begin{subequations}
\label{ch5_eq2}
\begin{align}
\label{expliciteqa}
\frac{\partial R_k(t,t_w)}{\partial t} &=\delta(t-t_w)-K_1R_k(t,t_w) \nonumber\\
&+\int_{t_w}^t \d s \Sigma_k(t,s)R_k(s,t_w)\\
\label{expliciteqb}
\frac{\partial C_k(t,t_w)}{\partial t} &= -K_1C_k(t,t_w)+\int_0^{t_w}\d s D_k(t,s)R_k(t_w,s) \nonumber\\
&+\int_{0}^t \d s \Sigma_k(t,s)C_k(s,t_w)
\end{align}
\end{subequations}
with the expressions of $D_k$ and $\Sigma_k$:
\begin{subequations}
\begin{align}
\label{selfenergy}
D_k(t,t_w) &= \frac{2k_BT}{D_L}\rho_k(t)\delta(t-t_w) \nonumber\\
&+\frac{K_2^2}{2} \int_{\bf q}\mathcal{V}_{k,q}^2C_q(t,t_w)C_{k-q}(t,t_w) \\
\Sigma_k(t,t_w) &= K_2^2\int_{\bf q} \mathcal{V}_{k,q}^2R_q(t,t_w)C_{k-q}(t,t_w).
\end{align}
\end{subequations}
The contribution from the first term in $D_k$ vanishes due to causality.

Defining the input quantities $K_1$ and $\mathcal{V}_{k,q}$ in equations (\ref{expliciteqa}) and (\ref{expliciteqb}) for the case of a quench is non-trivial. To gain some insight about these parameters, it is useful to compare the derivation with the treatment of Zaccarelli \textit{et al.} \cite{zaccarelli02}. The vertex term $\mathcal{V}_{k,q}$ in (\ref{ch5_eq1}) and (\ref{ch5_eq2}) involves the ``residual interactions'' in \cite{zaccarelli02}. Our definition of quench is an abrupt increase in the interaction strength, implying that $\mathcal{V}_{k,q}$ should be evaluated at the final parameter value. The variable $K_1$ contains the equal time density correlator. For an aging system, this must be evaluated at each instant of time since we are dealing with a non-stationary state due to the evolution of the system towards the equilibrium state at final parameter values. To determine $K_1$ we insist, as in \cite{cugliandolo93}, that for $\tau=(t-t_w) \ll t_w$ Eq.
(\ref{ch5_eq2}) obeys time-translation invariance and the FDR. Skipping some algebra, this condition will lead to
\begin{align}
\label{k1oft}
K_1(t)S_k&=TR_k(0)+K_2^2\int_0^t \int_{\bf q} \mathcal{V}_{k,q}^2C_{k-q}(t,s)\nonumber\\
\times& \bigg[\frac{1}{2}C_q(t,s)R_k(t,s)+R_q(t,s)C_k(t,s)\bigg]\d s.
\end{align}

Having derived the equations of motion for the two point correlators, we now proceed to calculate the corresponding susceptibilities for an aging structural glass. These susceptibilities are not exactly the same as, but related to, the three-point density correlators. Instead of attempting a direct calculation of the three-point correlators, the calculation of the susceptibilities is much easier and gives similar information. Let us impose an external potential $u^{ext}(\mathbf{r})$ that couples to one density; let the free energy functional in the presence of the potential be denoted by $F^u$. The equations of hydrodynamics for the density and the momenta are
\begin{equation}
\frac{\partial \rho}{\partial t}+\nabla\cdot(\rho{\bf v})=0,
\end{equation}
and
\begin{equation}
\frac{\partial\rho{\bf v}}{\partial t}=\eta\bigtriangledown^2{\bf v}+(\zeta+\eta/3)\nabla(\nabla\cdot{\bf v})-\rho\nabla\frac{\delta F^u}{\delta\rho} +\mathbf{\xi}(\mathbf{r},t).
\end{equation}
As is done in the previous case, it is possible to combine these two equations and write down the equation of motion for the density fluctuation alone as
\begin{equation}
\frac{\partial^2\delta\rho(t)}{\partial t^2}=D_L\bigtriangledown^2\frac{\partial\delta\rho}{\partial t}+\nabla\cdot(\rho\nabla\frac{\delta F^u}{\delta\rho})+\tilde{f}(\mathbf{r},t)
\end{equation}
where $D_L=(\zeta+4/3 \eta)/\rho_0$.

The modified RY free energy functional in the presence of the external potential will be given as \cite{saroj11}
\begin{align}
\beta F^u&=\int_{\bf r}\left[\rho({\bf r},t)\ln\left(\frac{\rho({\bf r},t)}{\rho_0}\right)-\delta\rho({\bf r},t)\right] \nonumber\\
&-\frac{1}{2}\int_{{\bf r},{\bf r}'}c(\mathbf{r}-\mathbf{r}')\delta\rho({\bf r},t)\delta\rho({\bf r}',t)+\beta\int_{\bf r}u^{{ext}}({\bf r})\delta\rho({\bf r},t)
\end{align}
where $\delta\rho({\bf r},t)=\rho({\bf r},t)-\rho_0$. Let us define the equilibrium static density $m({\bf r})$, satisfying
\begin{equation}
\beta\frac{\delta F^u}{\delta\rho({\bf r})}\bigg|_{\rho(\mathbf{r})=m(\mathbf{r})}=0
\end{equation}
and therefore, we will have
\begin{align}
&\ln\frac{m({\bf r})}{\rho_0}-\int_{{\bf r}'}c(\mathbf{r}-\mathbf{r}')\delta\rho({\bf r}')+\beta u^{{ext}}({\bf r})=0,\nonumber\\
&\ln\frac{m({\bf r})}{\rho_0}-\int_{{\bf r}'}c(\mathbf{r}-\mathbf{r}')[m({\bf r}')-\rho_0]+\beta u^{{ext}}({\bf r})=0.
\end{align}
Now we need the force density $\rho\nabla[{\delta F^u}/{\delta\rho({\bf r})}]$. For this purpose, we take the gradient of the above equation, remembering that $\nabla\int_{{\bf r}'}c(\mathbf{r}-\mathbf{r}')\rho_0$ is zero, since $\int_{{\bf r}'}c(\mathbf{r}-\mathbf{r}')\rho_0$ is independent of ${\bf r}$ because of the translational invariance of $c(\mathbf{r}-\mathbf{r}')$. Then 
\begin{equation}
\label{gradex}
\frac{\nabla m({\bf r})}{m({\bf r})}-\nabla\int_{{\bf r}'}c(\mathbf{r}-\mathbf{r}')m({\bf r}')+\beta\nabla u^{{ext}}=0.
\end{equation}
The fluctuation is taken around the equilibrium density $m({\bf r})$, which is inhomogeneous due to the presence of the external potential, and therefore, the total density at a point ${\bf r}$ is given by
\begin{equation}
\rho({\bf r},t)=m({\bf r})+\delta\rho({\bf r},t)
\end{equation}
and the force density is given as
\begin{widetext}
\begin{align}
(m({\bf r})+&\delta\rho({\bf r}))\nabla\frac{\beta\delta F^u}{\delta\rho({\bf r})}=\rho({\bf r})\nabla\bigg[\ln\left(\frac{\rho({\bf r})}{\rho_0}\right) 
-\int_{{\bf r}'}c(\mathbf{r}-\mathbf{r}')(\rho({\bf r}')-\rho_0)+\beta\nabla u^{{ext}}\bigg]\nonumber\\
&=\nabla\rho({\bf r})-(m({\bf r})+\delta\rho({\bf r}))\nabla\int_{{\bf r}'}c(\mathbf{r}-\mathbf{r}')(m({\bf r}')+\delta\rho({\bf r}'))
+(m({\bf r})+\delta\rho({\bf r}))\beta \nabla u^{{ext}}({\bf r})\nonumber\\
&=\nabla m({\bf r})+\nabla\delta\rho({\bf r})-m({\bf r})\nabla\int_{{\bf r}'}c(\mathbf{r}-\mathbf{r}')m({\bf r}')
-m({\bf r})\nabla\int_{{\bf r}'}c(\mathbf{r}-\mathbf{r}')\delta\rho({\bf r}')-\delta\rho({\bf r})\nabla\int_{{\bf r}'}c(\mathbf{r}-\mathbf{r}')m({\bf r}') \nonumber\\
&-\delta\rho({\bf r})\nabla\int_{{\bf r}'}c(\mathbf{r}-\mathbf{r}')\delta\rho({\bf r}')+\beta m({\bf r})\nabla u^{{ext}}({\bf r})
+\beta\delta\rho({\bf r})\nabla u^{{ext}}({\bf r})
\end{align}
Now, using Eq. (\ref{gradex}), the first, third and the seventh term will get cancelled. Also, from Eq. (\ref{gradex}),
\begin{equation}
-\delta\rho({\bf r})\nabla\int_{{\bf r}'}c(\mathbf{r}-\mathbf{r}')m({\bf r}')+\delta\rho({\bf r})\beta\nabla u^{{ext}}=-\delta\rho({\bf r})\frac{\nabla m({\bf r})}{m({\bf r})}.
\end{equation}
Using the above equation in the expression of the force density, we will have the final expression as
\begin{align}
\rho\nabla\frac{\delta\beta F^u}{\delta\rho({\bf r},t)}=\nabla\delta\rho({\bf r})-(m({\bf r})+\delta\rho({\bf r}))\nabla\int_{{\bf r}'}c(\mathbf{r}-\mathbf{r}')\delta\rho({\bf r}')-\delta\rho({\bf r})\frac{\nabla m({\bf r})}{m({\bf r})} .
\end{align}
Then, the time dependent force density is 
\begin{align}
\label{ch5_eq3}
\rho\nabla \frac{\delta \beta F^u}{\delta \rho({\bf r},t)}&=\nabla\int_{{\bf r}'}[\delta({\bf r}-{\bf r}')-\rho_0c(r-r')]\delta\rho({\bf r}',t)
-\nabla\int_{{\bf r}'}\delta m({\bf r})c(r-r')\delta\rho({\bf r}',t) -\delta\rho({\bf r},t)\nabla\int_{{\bf r}'}c(r-r')\delta\rho({\bf r}',t)\nonumber\\
&-\frac{\nabla m({\bf r})}{m({\bf r})}\int_{{\bf r}'}[\delta({\bf r}-{\bf r}')-m({\bf r})c(r-r')]\delta\rho({\bf r}',t)
\end{align}
where we have written the static inhomogeneous density $m(\mathbf{r})$ as the sum of two terms $\rho_0$, the homogeneous density in the absence of the external potential, and $\delta m({\bf r},t)$, the inhomogeneous density due to the external potential. We consider the case of weak perturbation by the external field: $\delta m({\bf r},t)$ is small.
Then we can linearize the force density equation by neglecting higher order terms in $\delta m({\bf r},t)$. The fourth term in the right hand side of Eq. (\ref{ch5_eq3}) will be modified as $\frac{\nabla \delta m({\bf r})}{\rho_0}\int_{{\bf r}'}[\delta({\bf r}-{\bf r}')-\rho_0c(r-r')]\delta\rho({\bf r}',t)$. Next we evaluate $\nabla \cdot \rho\nabla \frac{\delta \beta F^u}{\delta \rho({\bf r},t)}$ in $k$-space as
\begin{align}
\left[\nabla \cdot \rho\nabla \frac{\delta \beta F^u}{\delta \rho({\bf r},t)}\right]_{k}&=-k^2k_BT(1-\rho_0c_k)\delta\rho_{\bf k}(t)+k^2k_BT\int_{\bf q}\delta m_{{\bf k}-{\bf q}}c_q\delta\rho_{\bf q}(t)
+\frac{k_BT}{\rho_0}\int_{\bf q}\frac{{\bf k}\cdot({\bf k}-{\bf q})}{S_q}\delta m_{\bf k-\bf q}\delta\rho_{\bf q}(t)\nonumber\\
&+\frac{k_BT}{2}\int_{\bf q}{\bf k}\cdot[{\bf q}c_q+({\bf k-\bf q})c_{k-q}]\delta\rho_{\bf q}(t)\delta\rho_{\bf k-q}(t)
\end{align}
In the notation of Ref. \cite{biroli06}, we are interested in the $q\to0$ limit \cite{saroj12}. Let us consider the limit of a constant external potential that will produce a constant background density. Thus, $\delta m_k$ will be sharply localised at $k=0$ with a strength $\delta m_0$. 
Therefore, for this particular choice of the external perturbing field, we will have
\begin{align}
\left[\nabla \cdot \rho\nabla \frac{\delta \beta F^u}{\delta \rho({\bf r},t)}\right]_{k}=-\frac{k^2k_BT}{S_k}\delta\rho_{\bf k}(t)+k^2k_BT\delta m_{0}c_k\delta\rho_{\bf k}(t)
+\frac{k_BT}{2}\int_{\bf q}{\bf k}\cdot[{\bf q}c_q+({\bf k-\bf q})c_{k-q}]\delta\rho_{\bf q}(t)\delta\rho_{\bf k-q}(t)
\end{align}
Ignoring inertia and using the above form for the free-energy density, the equation of motion for the density fluctuation in Fourier space is
\begin{align}
D_Lk^2\frac{\partial\delta\rho_{\bf k}(t)}{\partial t}+\frac{k_BTk^2}{S_k}\delta\rho_{\bf k}(t)-k^2k_BT\delta m_{0}c_k\delta\rho_{\bf k}(t)
=\frac{k_BT}{2}\int_{\bf q}{\bf k}\cdot[{\bf q}c_q+({\bf k-\bf q})c_{k-q}]\delta\rho_{\bf q}(t)\delta\rho_{\bf k-q}(t)+\tilde{f}_k(t).
\end{align}
Let us divide the whole equation by $D_Lk^2$ and write $k_BT/D_LS_k$ as $K_1$ and $k_BT/D_Lk^2$ as $K_2$. For capturing the aging dynamics, however, we need to evaluate $K_1$ at each time step as we have explained in the calculation of the two-point correlator above. Therefore, we will have from the above equation
\begin{align}
\label{deltarhoeq}
\frac{\partial\delta\rho_{\bf k}(t)}{\partial t}+K_1(t)\delta\rho_{\bf k}(t)-\frac{k_BT\delta m_{0}c_k}{D_L}\delta\rho_{\bf k}(t)
=\frac{K_2}{2}\int_{\bf q}\mathcal{V}_{k,q}\delta\rho_{\bf q}(t)\delta\rho_{\bf k-q}(t)+{f}_k(t)
\end{align}
where we have written the vertex as $\mathcal{V}_{k,q}={\bf k}\cdot[{\bf q}c_\mathbf{q}+({\bf k-\bf q})c_{\mathbf{k}-\mathbf{q}}]$. The noise statistics of the bare noise $f_k(t)$ is as before in Eq. (\ref{noisestatistics}). Once we reach Eq. (\ref{deltarhoeq}), we use the diagrammatic perturbation calculation to obtain the equations of motion for the two-point correlators in the presence of the external field.

In this case, the bare propagator $R_{0k}$ is modified to
\begin{equation}
R_{0k}^{-1}=\frac{\partial}{\partial t}+K_1(t)-\frac{k_BT\delta m_{0}c_k}{D_L},
\end{equation}
and the rest of the calculation is same leading to the equations of motion for the two-point correlators denoted with a tilde on them to emphasize that they are evaluated in the presence of the external potential:
\begin{align}
\frac{\partial \tilde{R}_k(t,t_w)}{\partial t} =& -K_1(t)\tilde{R}_k(t,t_w) +\frac{k_BT\delta m_{0}c_k}{D_L}\tilde{R}_k(t,t_w) +\delta(t-t_w)
 +\int_{t_w}^t \d s \tilde{\Sigma}_k(t,s)\tilde{R}_k(s,t_w) \nonumber\\
\frac{\partial \tilde{C}_k(t,t_w)}{\partial t} =& -K_1(t)\tilde{C}_k(t,t_w) +\frac{k_BT\delta m_{0}c_k}{D_L}\tilde{C}_k(t,t_w) 
+\int_0^{t_w}\d s \tilde{D}_k(t,s)\tilde{R}_k(t_w,s) 
+\int_{0}^t \d s \tilde{\Sigma}_k(t,s)\tilde{C}_k(s,t_w),
\end{align}
where the expressions of $\tilde{D}_k$ and $\tilde{\Sigma}_k$ are given as
\begin{align}
\tilde{\Sigma}_k(t,t') &= \left(\frac{k_BT}{D_Lk^2}\right)^2\int_{\bf q} \mathcal{V}_{k,q}^2\tilde{R}_q(t,t')\tilde{C}_{k-q}(t,t') \\
\tilde{D}_k(t,t') &= \frac{2k_BT}{D_L}\rho_k(t)\delta(t-t')+ \frac{1}{2} \left(\frac{k_BT}{D_Lk^2}\right)^2 \int_{\bf q} \mathcal{V}_{k,q}^2\tilde{C}_q(t,t')\tilde{C}_{k-q}(t,t') \nonumber\\
&=\frac{2k_BT}{D_L}\rho_k(t)\delta(t-t')+\tilde{M}_k(t,t').
\end{align}

The equations of motion for the susceptibilities $\chi^C_k(t,t_w) = {\partial \tilde{C}_k(t,t_w)}/{\partial \delta m_0}|_{\delta m_0\to 0}$ and $\chi^R_k(t,t_w)= {\partial \tilde{R}_k(t,t_w)}/{\partial \delta m_0}|_{\delta m_0\to0}$, are given as
\begin{align}
 \label{chiRCa}
\frac{\partial \chi_k^{R}(t,t_w)}{\partial t} =& -K_1(t)\chi_k^{R}(t,t_w) + \int_{t_w}^t \d s {\Sigma}_k(t,s)\chi^{R}_k(s,t_w) 
+\int_{t_w}^t \d s \tilde{\Sigma}'_k(t,s){R}_k(s,t_w)+\mathcal{S}^R_k(t,t_w),\\ 
\label{chiRCb}
\frac{\partial \chi^{C}_k(t,t_w)}{\partial t} =& -K_1(t)\chi^{C}_k(t,t_w) + \int_0^{t_w}\d s {M}_k(t,s)\chi^{R}_k(t_w,s) 
+\int_0^{t_w}\d s  \tilde{M}'_k(t,s){R}_k(t_w,s) +\int_{0}^t \d s {\Sigma}_k(t,s)\chi^{C}_k(s,t_w) \nonumber \\
&+\int_{0}^t \d s \tilde{\Sigma}'_k(t,s){C}_k(s,t_w) +\mathcal{S}^C_k(t,t_w),
\end{align}
where $\tilde{\Sigma}'_k(t,s)=\partial \tilde{\Sigma}_k(t,s)/\partial\delta
m_0|_{\delta m_0\to0}$,  and $\tilde{M}'_k(t,s)={\partial
\tilde{M}_k(t,s)}/{\partial\delta m_0}|_{\delta m_0\to0}$. The expressions for the source terms $\mathcal{S}^R_k(t,t_w)$ and $\mathcal{S}^C_k(t,t_w)$ are
\begin{align}
\mathcal{S}_k^R(t,t_w)=\frac{k_BTc_k}{D_L}R_k(t,t_w)-\omega_k(t)R_k(t,t_w),
\nonumber\\
\mathcal{S}_k^C(t,t_w)=\frac{k_BTc_k}{D_L}C_k(t,t_w)-\omega_k(t)C_k(t,t_w)
\end{align}
where 
\begin{align}
\omega_k(t)=&\frac{K_2^2}{S_k}\int_0^t \int_{\bf q} \mathcal{V}_{k,q}^2\bigg[\chi^C_{k-q}(t,s)\bigg\{\frac{1}{2}C_q(t,s)R_k(t,s)+R_q(t,s)C_k(t,s)\bigg\} \nonumber\\
&+C_{k-q}(t,s)\bigg\{\frac{1}{2}\chi^C_q(t,s)R_k(t,s)+\frac{1}{2}C_q(t,s)\chi^R_k(t,s)
+\chi^R_q(t,s)C_k(t,s)+R_q(t,s)\chi_k^C(t,s)\bigg\}\bigg]\d s.
\end{align}

Now we need to solve these equations numerically to extract the predictions of the theory. However, a detailed solution of the full $k$-dependent equations requires huge computer time. Hence, we need to ``schematicise'' these equations to obtain a numerically tractable form. Simplified integral equations, keeping track of only the time dependence, has been extremely useful in extracting meaningful results from mode-coupling theory \cite{leutheusser84,kirkpatrick85,brader09} as it leads to a numerically manageable calculation. The schematic form of the two-point correlators in Eqs. \ref{ch5_eq2} will be
\begin{align}
\label{ch5_eq5}
\frac{\partial R(t,t_w)}{\partial t} =& \delta(t-t_w)-\mu(t) R(t,t_w) +4\lambda\int_{t_w}^t R(t,s)C(t,s)R(s,t_w)\d s \nonumber\\
\frac{\partial C(t,t_w)}{\partial t} =& -\mu(t) C(t,t_w)+ 2\lambda \int_0^{t_w}C^2(t,s)R(t_w,s) \d s 
+4\lambda\int_0^t C(t,s)R(t,s)C(s,t_w)\d s, 
\end{align}
where $\lambda$ is the interaction strength, $C(t,t_w)$ and $R(t,t_w)$ are the schematic forms of $C_k(t,t_w)$ and $R_k(t,t_w)$ respectively and $\mu(t)$ is the schematic version of $K_1(t)$:
\begin{equation}
\mu(t)=T+6\lambda\int_0^t C^2(t,s)R(t,s)\d s.
\end{equation} 

The schematic versions of $\chi^{C}_k(t,t_w)$ and $\chi^{R}_k(t,t_w)$ are written as $\chi_C(t,t_w)$ and $\chi_R(t,t_w)$ respectively. The final schematic forms of equations (\ref{chiRCa}) and (\ref{chiRCb}) will be
\begin{align}
\label{chi3schematic}
\frac{\partial \chi_R(t,t_w)}{\partial t} &+\mu(t)\chi_R(t,t_w)=4\lambda\int_{t_w}^t R(t,s)C(t,s)\chi_R(s,t_w)\d s 
+4\lambda\int_{t_w}^t R(t,s)\chi_C(t,s)R(s,t_w)\d s \nonumber\\
&+4\lambda\int_{t_w}^t \chi_R(t,s)C(t,s)R(s,t_w)\d s +\mathcal{S}_R(t,t_w) \\
\frac{\partial \chi_C(t,t_w)}{\partial t} & +\mu(t)\chi_C(t,t_w)= 4\lambda\int_0^{t_w}C(t,s)\chi_C(t,s)R(t_w,s)\d s 
+2\lambda\int_0^{t_w} C^2(t,s)\chi_R(t_w,s)\d s  \nonumber\\
&+4\lambda\int_0^{t} C(t,s)R(t,s)\chi_C(s,t_w)\d s  
+4\lambda\int_0^{t} \chi_C(t,s)R(t,s)C(s,t_w)\d s \nonumber\\ 
&+4\lambda\int_0^{t} C(t,s)\chi_R(t,s)C(s,t_w)\d s +\mathcal{S}_C(t,t_w)
\end{align}
with the source terms given as $\mathcal{S}_R(t,t_w)=[1-\omega(t)]R(t,t_w)$ and $\mathcal{S}_C(t,t_w)=[1-\omega(t)]C(t,t_w)$ where $\omega(t)$, the schematic form of $\omega_k(t)$, is given as
\begin{align}
\label{long_omega}
\omega(t)=12\lambda \int_0^t C(t,s)\chi_C(t,s)R(t,s)\d s +6\lambda\int_0^t C^2(t,s)\chi_R(t,s)\d s. 
\end{align}
\end{widetext}
It is also possible to obtain these wave vector-free equations of motion from a different approach, starting from a fully schematic version of the Langevin equation for the density fluctuation. We will outline the details of that calculation below.

\section{The schematic calculation for the growth kinetics in structural glasses}
\label{schematic}
In a schematic description one can throw away all the wave vectors and write Eq. (\ref{ch5_eq1}) as
\begin{equation}
\label{ch5_eq6}
\partial_t \phi(t)+\mu(t)\phi(t)=-\frac{g}{2}\phi^2(t)+f(t)
\end{equation}
where the information of the interaction strength goes into $g$ and we allow the frequency term $\mu(t)$ to be time-dependent that is appropriate for an aging system. Such an equation can also be obtained from a toy Hamiltonian $H=\frac{\mu(t)}{2}\phi^2(t)+\frac{g}{3!}\phi^3(t)$. $f(t)$ is a Gaussian white noise: $\langle f(t)f(t')\rangle=2T\delta(t-t')$. Once we have the Langevin equation for the density fluctuation, Eq. (\ref{ch5_eq6}), we can write down its perturbation expansion and obtain the equation of motion for the correlation function $C(t,t_w)=\langle \phi(t)\phi(t_w)\rangle$ and the response function $R(t,t_w)=\langle \partial \phi(t)/\partial f(t_w)\rangle$ in the same way as we did for the $k$-dependent case. Writing $g^2=4\lambda$, we will have the equations of motion for the response and correlation functions as
\begin{widetext}
\begin{align}
\label{ch5_eq7}
\frac{\partial R(t,t_w)}{\partial t} &= -\mu(t) R(t,t_w) +\delta(t-t_w)+4\lambda \int_{t_w}^t R(t,s)C(t,s)R(s,t_w)\d s \nonumber\\
\frac{\partial C(t,t_w)}{\partial t} &= -\mu(t) C(t,t_w)+2T R(t_w,t)+2\lambda\int_0^{t_w} C^2(t,s)R(t_w,s)\d s 
+4\lambda \int_0^t R(t,s)C(t,s)C(s,t_w)\d s.
\end{align}
\end{widetext}
In the equations of motion for $C(t,t_w)$, the second term in the right hand side, $2T R(t_w,t)$ will drop out because of the boundary condition on the response function. Note that these equations are exactly same as the schematic form of the full $k$-dependent equations for the two-point correlator Eqs. (\ref{ch5_eq5}). $\mu(t)$ can be obtained through a similar condition as used for the $k$-dependent case
\begin{equation}
\label{mueq}
\mu(t)=T+6\lambda\int_0^t C^2(t,s)\frac{\partial F(t,s)}{\partial s}\d s.
\end{equation}
These equations were also obtained by Franz and Hertz \cite{franz95} for the Amit-Roginsky model \cite{amit79}.

For a derivation of the three-point correlation functions through the schematic MCT approach, we need to impose an external field that couples to {\em two} fields at the same time. In the full $k$-dependent calculation of the equations, the presence of the external field that couples to one field will contribute a term $\int \epsilon(r)\delta\rho(r)$ in the free-energy functional $\mathcal{F}$. The force density is given by $-\rho\nabla[\delta \mathcal{F}/\delta \rho]$ and that will bring in a term that is linear in the field like $\epsilon_0\delta\rho$ for a constant field. To imitate this equation in the schematic approach, we must add a term that is quadratic in $\phi$ in the Hamiltonian:
\begin{equation}
H=\frac{\mu(t)}{2}\phi^2(t)+\frac{g}{3!}\phi^3(t)-\frac{\epsilon}{2}\phi^2(t).
\end{equation}

The $\epsilon$ term has the form of a shift in the frequency term $\mu(t)$: the Hamiltonian retains its form but $\mu\to\mu-\epsilon$. Then the calculation of the two-point correlation functions becomes same as before, and the equations for the correlation and response functions can be readily obtained from Eqs. (\ref{ch5_eq7}) with $\mu(t)$ being replaced by $\mu(t)-\epsilon$. We denote the response and correlation functions with a tilde to emphasize that they are evaluated in the presence of external field:
\begin{align}
\label{ch5_eq8}
\frac{\partial \tilde{R}(t,t_w)}{\partial t} &= -\delta(t-t_w)-\mu(t) \tilde{R}(t,t_w)+\epsilon \tilde{R}(t,t_w) \nonumber \\
&+4\lambda\int_{t_w}^t \tilde{R}(t,s)\tilde{C}(t,s)\tilde{R}(s,t_w)\d s \nonumber\\
\frac{\partial \tilde{C}(t,t_w)}{\partial t} &= -\mu(t) \tilde{C}(t,t_w)+\epsilon \tilde{C}(t,t_w) \nonumber\\
&+ 2\lambda \int_0^{t_w}\tilde{C}^2(t,s)\tilde{R}(t_w,s) \d s \nonumber \\
&+4\lambda\int_0^t \tilde{C}(t,s)\tilde{R}(t,s)\tilde{C}(s,t_w)\d s.
\end{align}
As before, we define the susceptibilities for the schematic case as
\begin{align}
\chi_C(t,t_w) &= \frac{\partial C(t,t_w)}{\partial \epsilon}\bigg|_{\epsilon=0} \nonumber\\
\chi_R(t,t_w) &= \frac{\partial R(t,t_w)}{\partial \epsilon}\bigg|_{\epsilon=0}.
\end{align}

\begin{widetext}
Then the equations of motion for the susceptibilities will be readily obtained from Eq. (\ref{ch5_eq8}).
\begin{align}
\label{ch5_eq9}
\frac{\partial \chi_R(t,t_w)}{\partial t} &+\mu(t)\chi_R(t,t_w)=4\lambda\int_{t_w}^t R(t,s)C(t,s)\chi_R(s,t_w)\d s 
+4\lambda\int_{t_w}^t R(t,s)\chi_C(t,s)R(s,t_w)\d s \nonumber \\
&+4\lambda\int_{t_w}^t \chi_R(t,s)C(t,s)R(s,t_w)\d s +\mathcal{S}_R(t,t_w) \nonumber\\
\frac{\partial \chi_C(t,t_w)}{\partial t} & +\mu(t)\chi_C(t,t_w)= 4\lambda\int_0^{t_w}C(t,s)\chi_C(t,s)R(t_w,s)\d s 
+2\lambda\int_0^{t_w} C^2(t,s)\chi_R(t_w,s)\d s \nonumber\\
&+4\lambda\int_0^{t} C(t,s)R(t,s)\chi_C(s,t_w)\d s 
+4\lambda\int_0^{t} \chi_C(t,s)R(t,s)C(s,t_w)\d s \nonumber\\
&+4\lambda\int_0^{t} C(t,s)\chi_R(t,s)C(s,t_w)\d s +\mathcal{S}_C(t,t_w),
\end{align}
with
\begin{align}
\label{ch5_eq10}
\mathcal{S}_R(t,t_w)=&[1-\omega(t)]R(t,t_w) \,\,\, {and}\,\,\,\mathcal{S}_C(t,t_w)=[1-\omega(t)]C(t,t_w),\nonumber\\
\mu(t)=&T+6\lambda\int_0^t C^2(t,s)R(t,s)\d s, \nonumber\\
\omega(t)=&12\lambda \int_0^t C(t,s)\chi_C(t,s)R(t,s)\d s+6\lambda\int_0^t C^2(t,s) \chi_R(t,s)\d s.
\end{align}
\end{widetext}
These equations are same as the schematic version of the full $k$-dependent equations of motion for the susceptibilities derived earlier. We solve these schematic equations (\ref{ch5_eq7}) and (\ref{ch5_eq9}) along with the definitions (\ref{ch5_eq10}) to obtain the growth kinetics of glassy correlations as we will discuss below.

\begin{figure}
\begin{center}
\includegraphics[width=8.6cm]{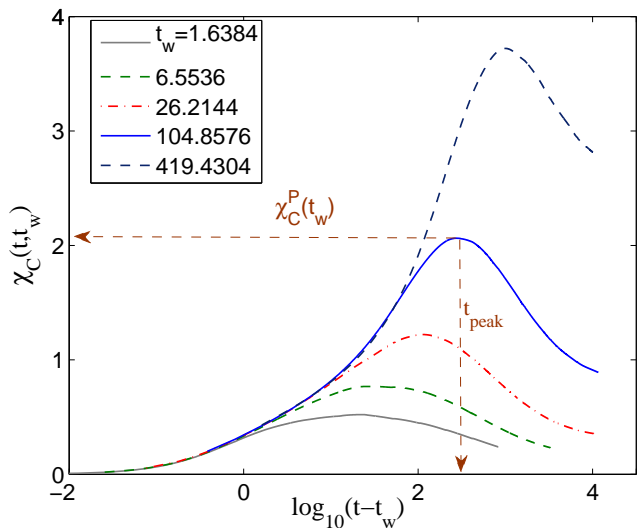}
\end{center}
\caption{The characteristic nonmonotonic decay of the three-point correlation function in an aging structural
glassy system, within schematic mode-coupling theory. $\chi_C^P(t_w)$, the peak value of $\chi_C(t,t_w)$ grows and shifts to higher $t_{peak}$ with increasing waiting time $t_w$. The final quenched parameter values are $T=1.0$ and $\lambda=2.01$.}
\label{largelambda}
\end{figure} 

\section{Results}
\label{coarseningresults}
The complicated algebraic details of the numerical solution of the equations governing growth kinetics of glassy correlations can be found in \cite{mythesis}. The numerical algorithm that we use here was first developed by Kim and Latz \cite{kim01,herzbach00} for the solution of the aging equations of the $p$-spin spherical spin glass model. Here we briefly describe the basic steps of this numerical algorithm and the interested reader is referred to \cite{herzbach00,mythesis} for the details.

The first step towards this solution is to write down the equations in terms of the correlation function and the integrated response function $F(t,t_w)$, defined as,
\begin{equation}
F(t,t_w)=-\int_{t_w}^t R(t,s)\d s.
\end{equation}
This transformation is advantageous since variation of $F$ is much smoother than that of the response function itself. Next we parametrize the equations in terms of $(t,\tau=t-t_w)$ instead of $(t,t_w)$ as in the original equations. This transformation is necessary since the decay of the correlation function is fast when the time difference $\tau=(t-t_w)$ is small and we need to solve the equations for very large time because the decay of the function is quite slow when $\tau$ is large. Therefore, we need to use the method of adaptive integration for the numerical solution. 
It requires a varying time-grid, which needs to be very small at short $\tau$ and large at large $\tau$ to resolve the full dynamics. If we use the $(t,t_w)$ parametrization we will have large time-grid for large $t_w$ even when $\tau=t-t_w$ is small, hence, we can not resolve the short time dynamics. This problem can be avoided in $(t,\tau)$ parametrization. The equations of motion for $F(t,\tau)$ and $C(t,\tau)$ will be
\begin{widetext}
\begin{align}
\label{ch5_schematic_twopoint}
(\partial_t+\partial_\tau)F(t,\tau) &= -1-\mu(t) F(t,\tau)-4\lambda\int_0^\tau \d s \frac{\partial F(t,s)}{\partial s}C(t,s)F(t-s,\tau-s) \nonumber \\
(\partial_t+\partial_\tau)C(t,\tau) &= -\mu(t) C(t,\tau)+2\lambda\int_\tau^t \frac{\partial C^2(t,s)}{\partial s} F(t-\tau,s-\tau)\d s 
-2\lambda C^2(t,t)F(t-\tau,t-\tau) \nonumber\\
&-4\lambda\int_\tau^t C(t,s)\frac{\partial F(t,s)}{\partial s}C(t-\tau,s-\tau)\d s 
-4\lambda\int_0^\tau C(t,s)\frac{\partial F(t,s)}{\partial s}C(t-s,\tau-s)\d s,
\end{align}
\end{widetext}
with 
\begin{equation}
\mu(t)=T-6\lambda\int_0^t C^2(t,s)\frac{\partial F(t,s)}{\partial s}\d s.
\end{equation}

We set temperature $T$ to unity and start with the initial conditions corresponding to a low density (or high temperature) liquid and quench the system into the glassy regime by setting the value of $\lambda$ to a large value and solve the equations in time. We see in Fig. \ref{correlationfn} that the two-point correlation function shows aging. The final value of $\lambda$ in this case is 2.01. We can define a relaxation time $t_r$ from the decay of the two-point functions as the time when the function decays to $1/e$. The behaviour of $t_r$ with $t_w$, as shown in Fig. \ref{relaxationtime}, agrees well with the numerical experiment of Kob and Barrat \cite{kob97}. For large $t_w$, the curve can be fitted with an algebraic form $t_r\sim t_w^\alpha$ with $\alpha \approx 0.8$. 
If we scale time by $t_r$ and the two-point correlation function shows data collapse, the aging of the system is termed as ``simple''. Such a data collapse is indeed found (inset of Fig. \ref{correlationfn}). This collapse of data suggests that we can associate the dynamics of the system with an evolving effective temperature. However, as we will see from the behaviour of the three-point correlation function [Fig. \ref{scaling_fit}(a)], such an association is more non-trivial than suggested by the decay of the two-point correlation function.

\begin{figure}
\begin{center}
\includegraphics[width=8.6cm]{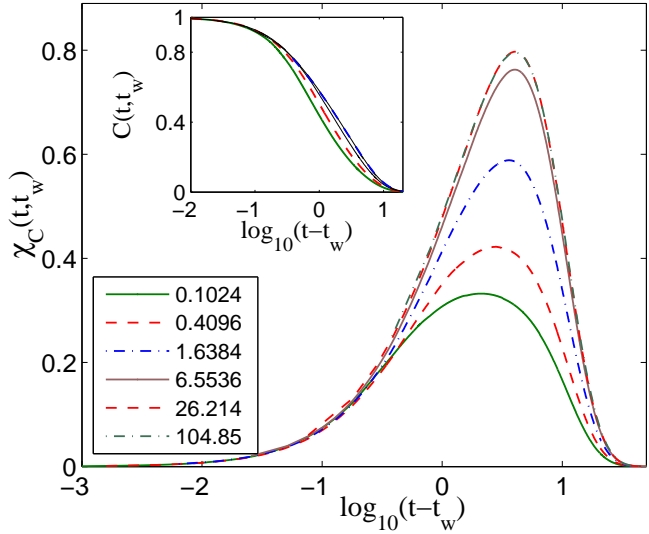}
\end{center}
\caption{When the quench is to liquid side, the three-point correlator grows to saturation. $\chi_C(t,t_w)$ is shown as a function of $t$ for various $t_w$ for $\lambda=0.75$. {\bf Inset:} The two-point correlator also progressively saturates for a quench in the liquid side. $C(t,t_w)$ as a function of $t$ for various $t_w$ is shown for the same parameter values.}
\label{lowlambda}
\end{figure}

Next we solve the equations of motion for the three-point correlators following a procedure similar to that used for the two-point functions. We define $\chi_F(t,t_w)=\partial \tilde{F}(t,t_w)/\partial \delta m_0|_{\delta m_0\to0}$. Then the equations of motion for the susceptibilities corresponding to the correlation function and the integrated response function in $(t,\tau)$ parametrization will be
\begin{widetext}
\label{ch5_eq11}
\begin{align}
(\partial_t+\partial_\tau)& \chi_F(t,\tau) = -\mu(t)\chi_F(t,\tau)-4\lambda\int_{0}^{\tau}\frac{\partial \chi_F(t,s)}{\partial s}C(t,s)F(t-s,\tau-s)\d s 
-4\lambda\int_0^{\tau}\frac{\partial F(t,s)}{\partial s}\chi_C(t,s)F(t-s,\tau-s)\d s \nonumber\\
&-4\lambda\int_0^{\tau}\frac{\partial F(t,s)}{\partial s}C(t,s)\chi_F(t-s,\tau-s)\d s+(1-\omega(t))F(t,\tau), \label{ch5_eq12} \\
(\partial_t+\partial_\tau)&\chi_C(t,\tau)= 1-\mu(t)\chi_C(t,\tau) -4\lambda\int_\tau^{t}C(t,s)\chi_C(t,s)\frac{\partial F(t-\tau,s-\tau)}{\partial s} \d s 
-2\lambda\int_\tau^{t} C^2(t,s)\frac{\partial \chi_F(t-\tau,s-\tau)}{\partial s}\d s \nonumber\\
&-4\lambda\int_\tau^t \chi_C(t,s)\frac{\partial F(t,s)}{\partial s}C(t-\tau,s-\tau)\d s 
-4\lambda\int_\tau^t C(t,s)\frac{\partial \chi_F(t,s)}{\partial s}C(t-\tau,s-\tau)\d s \nonumber\\
&-4\lambda\int_\tau^t C(t,s)\frac{\partial F(t,s)}{\partial s}\chi_C(t-\tau,s-\tau)\d s 
-4\lambda\int_0^\tau \chi_C(t,s)\frac{\partial F(t,s)}{\partial s}C(t-s,\tau-s)\d s \nonumber \\
&-4\lambda\int_0^\tau C(t,s)\frac{\partial \chi_F(t,s)}{\partial s}C(t-s,\tau-s)\d s 
-4\lambda\int_0^\tau C(t,s)\frac{\partial F(t,s)}{\partial s}\chi_C(t-s,\tau-s)\d s+(1-\omega(t))C(t,\tau) 
\label{ch5_eq13}
\end{align}
\end{widetext}
with $\mu(t)$ and $\omega(t)$ being
\begin{align}
\mu(t) &= T - 6\lambda \int_0^t C^2(t,s)\frac{\partial F(t,s)}{\partial s}\d s \nonumber \\
\omega(t)&= -12\lambda \int_0^t C(t,s)\chi_C(t,s)\frac{\partial F(t,s)}{\partial s}\d s \nonumber\\
&-6\lambda\int_0^t C^2(t,s)\frac{\partial \chi_F(t,s)}{\partial s}\d s
\end{align}

\begin{figure}
\begin{center}
\includegraphics[width=8.6cm]{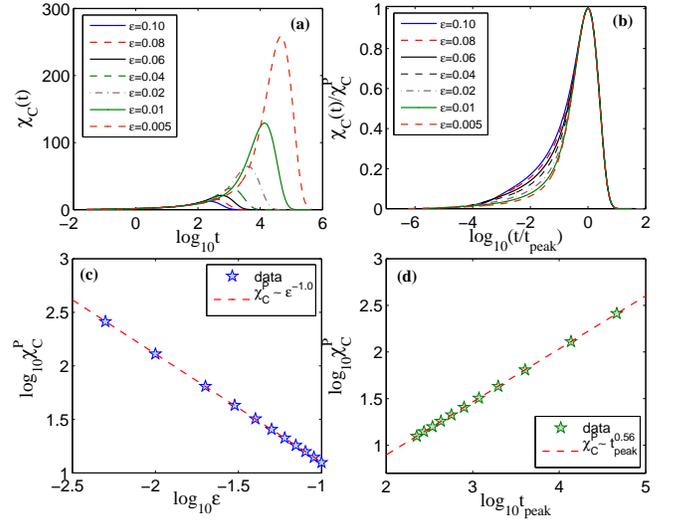}
\end{center}
\caption{The three-point correlation function and correlation volume in equilibrium. $\epsilon \equiv |\lambda-\lambda_c|$ defines the control parameter. (a) The characteristic nonmonotonic behaviour of $\chi_C(t)$ as a function of $t$ for various $\epsilon$. (b) Data collapse is obtained in the $\alpha$-regime when $\chi_C(t)$ is scaled with $\chi_C^P$ and time with $t_{peak}$. (c) Correlation volume $\chi_C^P \sim \epsilon^{-1.0}$. (d) $\chi_C^P\sim t_{peak}^{0.56}$. (Taken from Fig. 4 of Ref. \cite{saroj12})}
\label{ss_chi3scaling}
\end{figure}

The characteristic nonmonotonic behaviour of the three-point correlator and its dependence on $t_w$ is shown in Fig. \ref{largelambda}. For a fixed initial condition that corresponds to high-temperature liquid with negligible interactions, we examine how the three-point correlator evolves when the quench is to the liquid state or to the glassy state. The quench is defined by the value of $\lambda$. Let us call $\chi_C^P(t_w)$ the peak value of $\chi_C(t,t_w)$ and $t_{peak}$ the time at which $\chi_C(t,t_w)$ attains the peak. If $\lambda$ is still in the liquid phase, both $\chi_C^P$ and $t_{peak}$ increase and then saturate to a certain value depending on the quench (Fig. \ref{lowlambda}). 

If we take the value of $\lambda$ that corresponds to liquid state and the limit of waiting time $t_w$ to infinity, the system will reach equilibrium and time-translational invariance will be restored. Let us take eqs. (\ref{ch5_eq5})-(\ref{long_omega}), take the limit of $t_w\to\infty$ and impose FDR. The resulting equations of motion, describing an equilibrium system, are
\begin{subequations}
\begin{align}
\frac{\partial C(t)}{\partial t}&+TC(t)+2\lambda\int_0^tC(t-s)\frac{\partial C(s)}{\partial s} ds=0, \\
\frac{\partial \chi_C(t)}{\partial t} & +T\chi_C(t)+\int_0^t\Sigma(t-s)\frac{\partial C(s)}{\partial s} ds \nonumber\\
&+\int_0^t m(t-s)\frac{\partial\chi_C(s)}{\partial s} ds=C(t),
\end{align}
\end{subequations}
with the memory kernels $\Sigma(t)=4\lambda C(t)\chi_C(t)$ and $m(t)=2\lambda C^2(t)$. In obtaining these equations we have neglected $C(t\to\infty)$, but this is fine since we are in the liquid state. The behaviour of the equilibrium $\chi_C(t)$ as a function of $t$ is shown in Fig. \ref{ss_chi3scaling}(a). One obtains data collapse if time is scaled by $t_{peak}$ and $\chi_C(t)$ by $\chi_C^P$. This should not be surprising since the system is in equilibrium in this case. The peak value $\chi_C^P$, which is a measure of correlation volume, grows as $(\lambda-\lambda_c)^{-1}$ and $\chi_C^P\sim t_{peak}^{0.56}$ [Fig. \ref{ss_chi3scaling}(c) and (d)].
The equilibrium limit of our theory corresponds to the $\mathbf{q}\to 0$ limit of Ref. \cite{biroli06} and the results in these two limits of the corresponding theories do agree precisely \cite{saroj12}. Our theory can also be compared in a crude sense with \cite{karmakar09} and the predictions of our theory match with their findings. However, a more detailed comparison with \cite{biroli06} or, for that matter with \cite{karmakar09}, will require calculating the susceptibilities with respect to a spatially varying potential, a task we did not attempt in this work.

A completely different scenario arises for a quench into the glassy regime, the aging continues uninterrupted. In our notation, the mode-coupling transition occurs at $\lambda_c=2.0$. Let us now discuss two important characteristics of the dynamics. First, is there any difference in the aging scenario if we quench from two different high temperature states, say $\lambda_1$ and $\lambda_2$ with $\lambda_1<\lambda_2$, to the same low temperature state? After a certain waiting time, the system loses the information of the initial conditions and settles to the steady aging. The aging dynamics that started with the initial condition $\lambda_1$ will start following the aging dynamics that started with the initial condition $\lambda_2$ after a waiting time that depends on $\lambda_1$ \cite{kob97,kob00}. Thus, although the initial aging dynamics depends on the initial condition, after a suitable lapse of waiting time the dynamics is characterized by the final quenched parameter. However, the dynamics has more complicated dependence on various parameters and it can not be characterized by an evolving effective temperature as will be shown below. 
Second, within MCT, is there any difference between two different quenches characterized by $\lambda_{f1}$ and $\lambda_{f2}$ when both are greater than $\lambda_c$? The answer is nontrivial. The final quench parameter acts as the driving force of aging. Let us say $\lambda_{f2}>\lambda_{f1}$, then after the same waiting time, the second system will become more sluggish than the first one. In Fig. 2 of \cite{saroj12} we presented the growth dynamics for quench to $\lambda=2.00$ and here we present the growth dynamics for quench to $\lambda=2.01$ in Fig. \ref{largelambda}; the initial conditions are same in both the cases. A careful examination between these two figures reveals that for the same $t_w$, $\chi_C^P(\lambda=2.00)>\chi_C^P(2.01)$. It is important to note that the system at various waiting times can not be characterised by an evolving effective temperature. If it had been so, the system would not have been able to distinguish between $\lambda_{f1}$ and $\lambda_{f2}$ since the relaxation time becomes infinite once the system reaches the states corresponding to $\lambda_c$. But during the aging (or coarsening), the system seems to already have the information of its final quench parameter value.

In Fig. \ref{largelambda} we show the behaviour of $\chi_C(t,t_w)$ as a function of $t$ for various $t_w$ for a quench corresponding to $\lambda=2.01$. In this case $\chi_C^P(t_w)$ grows without bound as the waiting time increases with the growth law $\chi_C^P \sim t_w^a$ with the exponent $a \simeq 0.5$ [Fig. \ref{scaling_fit}(b)] in agreement with numerical experiments \cite{parisi99}. Since $\chi_C^P$ is the measure of an effective correlation volume, our theory quantifies the idea of {\it domain growth of glassy correlations} starting from a liquid background. 
$t_{peak}$ is a measure of the relaxation time of the system and $t_{peak}$ increases with the waiting time as $t_{peak}\ \sim t_w^b$ with $b\simeq 0.8$. This relation has not been tested in simulations. We have discussed about another relaxation time $t_r$, extracted from the two-point correlation function. It is found that these two relaxation times are related as $t_{peak}\simeq 4t_r^\beta$ with $\beta \approx 1.0$. 
It is important to note here that regardless of the detailed values of the various parameters, it is significant that our theory and also the simulations of \cite{parisi99,parsaeian08} obtained a very sublinear growth of the ``domain size''

\begin{figure*}
\begin{center}
\includegraphics[width=17cm]{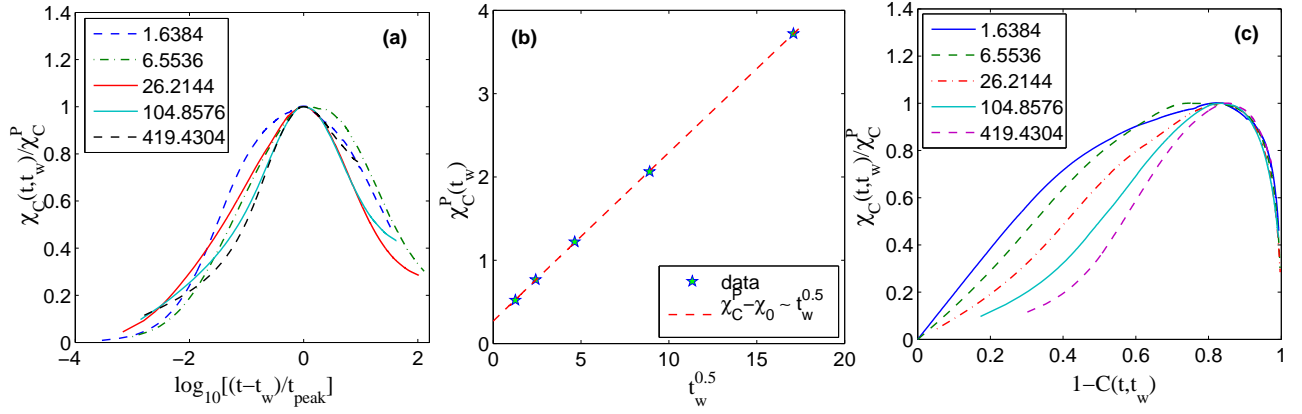}
\end{center}
\caption{(a) No data collapse is observed when $\chi_C(t,t_w)$ is scaled by $\chi_C^P(t_w)$ and time by $t_{peak}$: this imply that the behaviour of the three-point correlator at various $t_w$ can not be seen as the equilibrium dynamics at an evolving effective temperature. 
The behaviour of $\chi_C$ does not agree with the expectation of ``simple aging'' that was suggested by the data collapse obtained for the two-point function in Fig. \ref{correlationfn}. (b) The peak height $\chi_C^P(t_w)$ that is a measure of the correlation volume is proportional to $t_w^{1/2}$. (f) When we scale $\chi_C(t,t_w)$ by $\chi_C^P$ and plot it as a function of $1-C(t,t_w)$, data collapse is obtained in $\alpha$-relaxation regime.}
\label{scaling_fit}
\end{figure*} 

As we have discussed in \cite{saroj12} and also shown in Fig. \ref{scaling_fit}(a), we do not see any data collapse when we scale $\chi_C(t,t_w)$ by $\chi_C^P(t_w)$ and time by $t_{peak}$. 
We have also shown in \cite{saroj12} that we did not find any data collapse following the scaling relations suggested in \cite{parsaeian08}, this may be because we measured different quantities. As shown in Fig. \ref{scaling_fit}(c), if we scale $\chi_C(t,t_w)$ by $\chi_C^P$ and plot them as
a function of $1-C(t,t_w)$, we obtained data collapse in the $\alpha$-relaxation
regime. However, we do not understand the origin of this scaling. 
The lack of the scaling for the three-point correlator as shown in Fig. \ref{scaling_fit}(a) implies that describing the dynamics in terms of an evolving effective temperature misses some key points and the $t_w$-dependent properties we extract do not correspond to those of an equilibrium system at an evolving $\lambda$ or temperature \cite{footnote2}. This appears to contradict to what is suggested by the behaviour of the two-point correlation function as shown in the inset of Fig. \ref{correlationfn} or in simulation \cite{kob97} or mode-coupling theory \cite{sunil09}. However, it is possible that the monotonic decay of the two-point correlator masks the deviations or, more likely, the three-point function contains additional independent information and the nonmonotonic nature of the later is more sensitive to departures from ``simple aging''. We have seen that even if the quench is in the liquid state, the system didn't show any data collapse of the short described in Fig. \ref{scaling_fit}(a).

\section{Growth kinetics of $p$-spin spin glass mean-field model}
\label{spinglass}
There exists a deep connection between the dynamics of certain spin-glass models and that of the structural glasses \cite{kirkthiru87,kirkpatrick87}. Moreover, the dynamics of the mean-field $p$-spin spin-glass models can be treated analytically and that promises deeper understanding for the structural glasses as well \cite{castellani05,kim01}. Even though the aging dynamics in the mean-field models of spin glasses has been studied in detail \cite{cugliandolo93,cugliandolo94,kim01,sunil09}, the role played by the length scale obtained from the multi-point correlators and their characteristics in general have not been studied. In this section, we extend the calculation for mean-field $p$-spin spin-glass models to capture the growth kinetics upon abrupt quench.

Let us start with the microscopic Hamiltonian for the $p$-spin spherical spin-glass model
\begin{equation}
\mathcal{H}=-\sum_{i_1>i_2>\ldots >i_p=1}^N J_{i_1\ldots i_p}S_{i_1}S_{i_2}\ldots S_{i_p},
\end{equation}
where the couplings $J_{i_1\ldots i_p}$ are Gaussian random variables with zero mean and variance $p!/2N^{p-1}$. $S_i$'s are the spin variables obeying the spherical constraint $\sum_{i=1}^N S_i^2=1$. This model shows an equilibrium phase transition at a temperature $T_s$ when the system goes from the paramagnetic phase to the spin-glass state characterised by the one-step replica symmetry breaking \cite{kim01,castellani05}. $T_s$ is lower than $T_d$, the dynamic temperature when the system goes to non-ergodic state as also given by mode-coupling theory for structural glasses.

To write down the dynamics for this model, let us consider the Langevin equation
\begin{equation}
\frac{\partial S_i(t)}{\partial t}=-z(t)S_i(t)-\frac{\partial \mathcal{H}}{\partial S_i(t)}+\eta_i(t),
\end{equation}
where we have set the kinetic coefficient to unity and $z(t)$ is a Lagrange multiplier to satisfy the spherical constraint. $\eta_i(t)$ satisfies the white noise statistics. In the limit $N\to \infty$, one can characterize the system with a scalar variable \cite{stanley68}. In that limit one can treat the dynamics analytically through the standard dynamical field theory \cite{msr73,janssen76,dedominicis76,jensen81}. The calculation for the two-point correlation function is tedious, but quite standard and well-known in the literature \cite{kirkpatrick87,crisanti93,castellani05,kim01}, so let us just quote the results here.
\label{ch5_eq14}
\begin{align}
\frac{\partial C(t,t_w)}{\partial t} &= -z(t) C(t,t_w)+\frac{p}{2}\int_0^{t_w} C^{p-1}(t,s)R(t_w,s)\d s \nonumber\\
& +\frac{p(p-1)}{2} \int_0^t C^{p-2}(t,s)R(t,s)C(s,t_w)\d s, \\
\frac{\partial R(t,t_w)}{\partial t} &= -z(t) R(t,t_w) +\delta(t-t_w) \nonumber\\
&+\frac{p(p-1)}{2} \int_{t_w}^t C^{p-2}(t,s)R(t,s)R(s,t_w)\d s,
\end{align}
with
\begin{equation}
z(t)=T+\frac{p^2}{2}\int_0^t C^{p-1}(t,s)R(t,s)\d s.
\end{equation}

\begin{figure}
\begin{center}
\includegraphics[height=6cm]{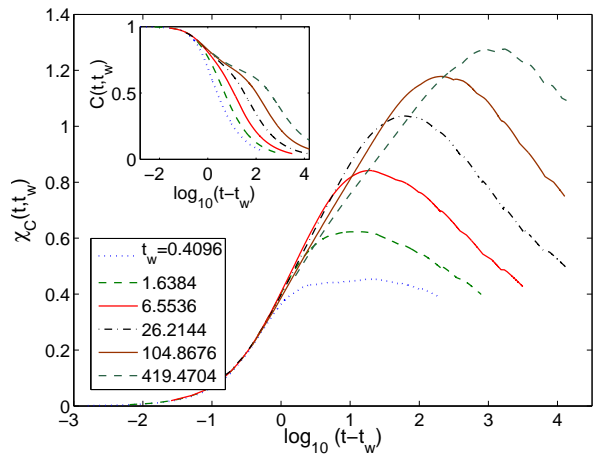}
\end{center}
\caption{Evolution of the three-point correlator for a quench corresponding to $T=0.5$ starting from a high temperature state of the mean-field $p$-spin spin-glass model. {\bf Inset:} The two-point correlation function as a function of $t$ for various $t_w$ for the same set of parameter values. Various $t_w$ are same as in the main figure.}
\label{spinglass_coarsening}
\end{figure}

It has been shown in Ref. \cite{berthier05} that susceptibilities, defined as the derivatives of the two-point correlator with respect to a number of control parameters like temperature, pressure or density, are capable of capturing information about the length scale of glassy dynamics \cite{berthier07a,berthier07b}. In particular, the derivatives of  $C$ and $R$ with respect to $T$ has the status of a three-point correlator. Thus, to capture the growth kinetics of the spin-glass system, we define non-stationary susceptibilities as follows:
\begin{align}
\chi_C(t,t_w)=-\frac{\partial C(t,t_w)}{\partial T} \nonumber \\
\chi_R(t,t_w)=-\frac{\partial R(t,t_w)}{\partial T}. \nonumber 
\end{align}
Then we will obtain the equations governing the domain growth for glassy regions as
\begin{widetext}
\begin{subequations}
\begin{align}
\frac{\partial \chi_C(t,t_w)}{\partial t} &= -z(t) \chi_C(t,t_w)+[1-\omega(t)]C(t,t_w) 
+\frac{p(p-1)}{2}\int_0^{t_w} C^{p-2}(t,s)\chi_C(t,s)R(t_w,s)\d s \nonumber\\
&+\frac{p}{2}\int_0^{t_w} C^{p-1}(t,s)\chi_R(t_w,s)\d s  
+\frac{p(p-1)(p-2)}{2} \int_0^t C^{p-3}(t,s)\chi_C(t,s)R(t,s)C(s,t_w)\d s \nonumber\\
&+\frac{p(p-1)}{2} \int_0^t C^{p-2}(t,s)\chi_R(t,s)C(s,t_w)\d s  
+\frac{p(p-1)}{2} \int_0^t C^{p-2}(t,s)R(t,s)\chi_C(s,t_w)\d s \\
\frac{\partial \chi_R(t,t_w)}{\partial t} &= -z(t) \chi_R(t,t_w)+[1-\omega(t)]R(t,t_w) 
+\frac{p(p-1)(p-2)}{2} \int_{t_w}^t C^{p-3}(t,s)\chi_C(t,s)R(t,s)R(s,t_w)\d s \nonumber\\
&+\frac{p(p-1)}{2} \int_{t_w}^t C^{p-2}(t,s)\chi_R(t,s)R(s,t_w)\d s 
+\frac{p(p-1)}{2} \int_{t_w}^t C^{p-2}(t,s)R(t,s)\chi_R(s,t_w)\d s,
\end{align}
\end{subequations}
with the definition of $\omega(t)$ as
\begin{equation}
\omega(t)=\frac{p^2(p-1)}{2}\int_0^tC^{p-2}(t,s)\chi_C(t,s)R(t,s)\d s
+\frac{p^2}{2}\int_0^tC^{p-1}(t,s)\chi_R(t,s)\d s.
\end{equation}

\end{widetext}

The numerical algorithm for solving these equations of motion are same as for the case of structural glasses (see \cite{herzbach00,mythesis} for details). Here we show the results for the particular case of $p=3$. We find that the correlation volume has a behaviour similar to the structural glasses; the peak value $\chi_C^P$ grows and shifts to larger $t-t_w$ as one waits longer in the quenched state (Fig. \ref{spinglass_coarsening}). However, the growth law is different in this case; for a quench to $T=0.5$, $\chi_C^P(t_w)\sim t_w^{0.13}$ which is much smaller than the exponent found for the growth of the correlation volume of structural glasses. More detailed study is required for a deeper understanding of the growth kinetics of spin-glasses.

\section{Aging under shear: cut-off of the growing length scale in structural glass}
\label{aging_shear}
Since the theoretical system size is infinite, the system will never equilibrate when we quench it from the liquid state to below $\lambda_c$, the MCT transition point; the length scale and relaxation time will keep on increasing as the waiting time $t_w$ increases. However, imposing a nonzero shear will force the system to reach a steady state cutting-off the growth of the length scale and the relaxation time $\tau$ of the system. Shearing is like stirring the system at a time scale $\gdot^{-1}$ where $\gdot$ is the shear rate and the system will be affected by shear only if $\tau \geq \gdot^{-1}$. In this section we explore the effect of shear on the coarsening dynamics of structural glasses.

\begin{figure*}
\begin{center}
\includegraphics[height=6cm]{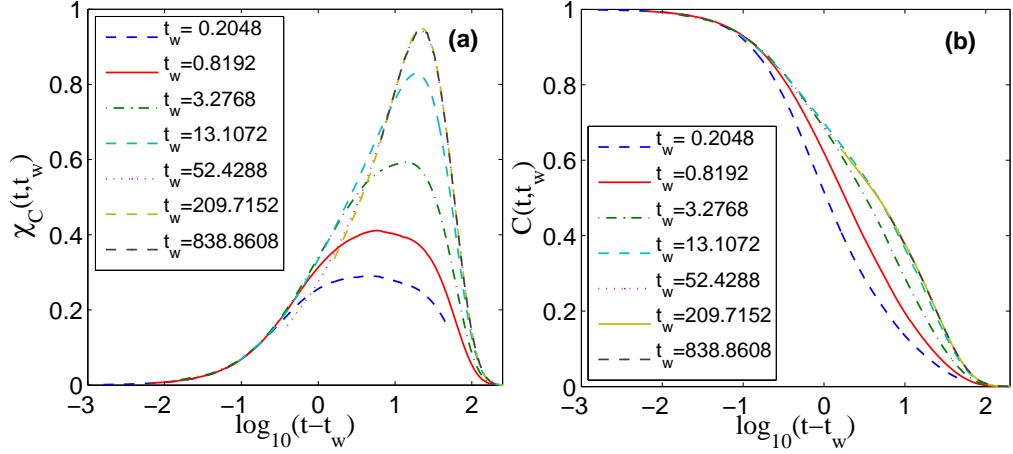}
\end{center}
\caption{(a) The growth of the three-point correlator $\chi_C(t,t_w)$ under aging gets cut-off by shear when the waiting time becomes of the order of inverse shear rate. The imposed shear rate is $\gdot=10^{-2}$. (b) The two-point correlator for the aging system under shear also reaches a steady state when $t_w\sim\gdot^{-1}$.}
\label{ch5_gdot_2}
\end{figure*}

Within mode-coupling theory, shear enters as two important effects \cite{fuchs03,brader07,brader08,chong09,kuni04,indrani95,kuni02}: first, shear modifies the equilibrium structure factor and reduces the structure height, second, it modifies the mode-coupling vertex and the MCT kernel loses its weight when the time scale becomes of the order of inverse shear rate $\dot{\gamma}^{-1}$. Even though both these modifications have their importances, for a qualitative understanding, it seems enough to keep the second effect in the theory \cite{fuchs03,kuni04}. Moreover, if we assume isotropy \cite{fuchs03} the structure factor of the fluid remains unchanged even under shear \cite{footnote1}. In that case, the critical input that shear brings into the calculation is the advection of wave vectors due to shear and the $\mathbf{k}\equiv \mathbf{k}(t)$ vectors at time $t$ gets coupled with the wave vectors $\mathbf{k}(t')$ at time $t'$ and rest of the calculation remains same as that for an unsheared fluid. The advected wave vector can be calculated for a shear in the $x$-direction and the velocity gradient in the $y$-direction as
\begin{equation}
\mathbf{k}(t)=\mathbf{k}(0)+\dot{\gamma}tk_x\hat{y}.
\end{equation}
We will not write this time index on the wave vectors explicitly, but instead use the notation that the associated wave vector of a quantity at time $t$ is also at that same time. For the two-time quantities like the correlation and response functions, we will assume the meaning of index $\mathbf{k}$ being the coupling of wave vector $\mathbf{k}(t)$ with $\mathbf{k}(t_w)$ and write the variables as $C_k(t,t_w)$ and $R_k(t,t_w)$ as a compact notation. Let us first calculate the equation of motion for the two-point correlator under shear. 

The equation of motion for the density fluctuation of a dense fluid can be obtained from the continuity equations (for the density and momenta) of hydrodynamics (see Eq. (\ref{ch5_eq1}) for the derivation):
\begin{align}
\frac{\partial \delta \rho_k(t)}{\partial t} +K_1(t)\delta\rho_k(t) =\frac{K_2}{2}\int_{\bf q} \mathcal{V}_{k,q}\delta\rho_q(t)\delta\rho_{k-q}(t)+\eta_k(t),
\end{align}
where $D_L=(\zeta+4\eta/3)/\rho_0$, $\mathcal{V}_{k,q}= \mathbf{k}\cdot[\mathbf{q}c_q+(\mathbf{k}-\mathbf{q})c_{k-q}]$ and the noise obeys the following statistics:
\begin{equation}
\langle \eta_k(t)\eta_{k'}(t')\rangle=\frac{2k_BT}{D_L}\rho_k(t)\delta(\mathbf{k}+\mathbf{k}')\delta(t-t').
\end{equation}

Then following a similar calculation as was done above in the case of unsheared fluid, keeping the advection of wave vector in mind, we will obtain the equations of motion for an aging fluid under shear as
\begin{subequations}
\begin{align}
\label{ch5_eq15}
\frac{\partial R_k(t,t_w)}{\partial t} =& -K_1(t)R_k(t,t_w)+\delta(t-t_w) \\
&+\int_{t_w}^t \d s \Sigma_k(t,s)R_k(s,t_w) \nonumber\\
\frac{\partial C_k(t,t_w)}{\partial t} =& -K_1(t)C_k(t,t_w)+\int_0^{t_w}\d s D_k(t,s)R_k(t_w,s) \nonumber\\
&+\int_{0}^t \d s \Sigma_k(t,s)C_k(s,t_w),
\end{align}
\end{subequations}

\begin{widetext}
with the expressions for $D_k$ and $\Sigma_k$ are
\begin{align}
\label{ch5_eq16}
D_k(t,t') &= \frac{2k_BT}{D_L}\rho_k(t)\delta(t-t') + \frac{1}{2} \left(\frac{k_BT}{D_Lk(t)k(t')}\right)^2 \int_{\bf q} \bigg(\mathbf{k}(t)\cdot[\mathbf{q}(t)c_{q(t)}+(\mathbf{k}(t)-\mathbf{q}(t))c_{k(t)-q(t)}]\bigg) \times \nonumber\\
&\bigg(\mathbf{k}(t')\cdot[\mathbf{q}(t')c_{q(t')}+(\mathbf{k}(t')-\mathbf{q}(t'))c_{k(t')-q(t')}]\bigg)
C_q(t,t')C_{k-q}(t,t') \nonumber\\
\Sigma_k(t,t') &= 
\left(\frac{k_BT}{D_Lk(t)k(t')}\right)^2 \int_{\bf q} \bigg(\mathbf{k}(t)\cdot[\mathbf{q}(t)c_{q(t)}+(\mathbf{k}(t)-\mathbf{q}(t))c_{k(t)-q(t)}]\bigg) \times \nonumber\\
&\bigg(\mathbf{k}(t')\cdot[\mathbf{q}(t')c_{q(t')}+(\mathbf{k}(t')-\mathbf{q}(t'))c_{k(t')-q(t')}]\bigg)
R_q(t,t')C_{k-q}(t,t').
\end{align}

Upon replacing $D_k$ and $\Sigma_k$ from Eq. (\ref{ch5_eq16}) into Eq. (\ref{ch5_eq15}) we will get the complete mode-coupling equation that we need to solve self consistently in order to see what the theory predicts for an aging system under shear. As before, since we are interested in the schematic limit, the detailed statistics of noise is not important.

The equations of motion for the growth kinetics are derived in the same procedure as before and we will just present the final equations here.
\begin{align}
\frac{\partial \chi_k^{R}(t,t_w)}{\partial t} =& -K_1(t)\chi_k^{R}(t,t_w)  +\int_{t_w}^t \d s {\Sigma}_k(t,s)\chi^{R}_k(s,t_w)  
+\int_{t_w}^t \d s\frac{\partial \tilde{M}_k(t,t_w)}{\partial \epsilon}\bigg|_{\epsilon\to0} R_k(s,t_w)+\left(\frac{k_BTc_k}{D_L}-\omega_k(t)\right){R}_k(t,t_w) \nonumber\\
\frac{\partial \chi^{C}_k(t,t_w)}{\partial t} =& -K_1(t)\chi^{C}_k(t,t_w)  +\int_0^{t_w}\d s {M}_k(t,s)\chi^{R}_k(t_w,s) 
+\int_{0}^t \d s {\Sigma}_k(t,s)\chi^{C}_k(s,t_w) +\int_0^{t_w}\d s\frac{\partial \tilde{M}_k(t,t_w)}{\partial \epsilon}\bigg|_{\epsilon\to0} R_k(s,t_w)  \nonumber \\
&+\int_0^t\d s \frac{\partial \tilde{\Sigma}_k(t,t_w)}{\partial \epsilon} \bigg|_{\epsilon\to0}C_k(s,t_w)
+\left(\frac{k_BTc_k}{D_L}-\omega_k(t)\right){C}_k(t,t_w),
\end{align}
where $K_1(t)$ and $\omega_k(t)$ have structure similar to that in the unsheared case and we will present their schematic form below.

\begin{figure*}
\begin{center}
\includegraphics[height=6cm]{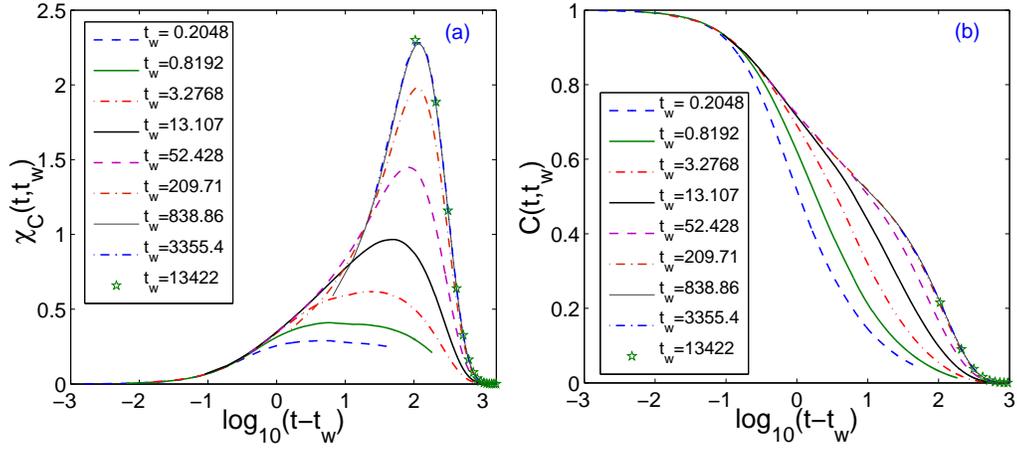}
\end{center}
\caption{When the system is quenched from high temperature liquid phase to deep in glassy phase with the quench characterised by the parameters $T=1.0$ and $\lambda=2.0$, we expect the system to show aging behaviour for ever. However shear (with $\gdot=10^{-3}$) cuts-off the aging behaviour and drives the system to its steady state after an waiting time $t_w\sim10^3$. (a) The behaviour of $\chi_C(t,t_w)$ as a function of $t$ for various $t_w$. (b) Behaviour of the two-point correlator $C(t,t_w)$.}
\label{gdot_3}
\end{figure*}

Now we will schematicise these equations. First, let us look at the memory kernel. Due to advection of wave vectors, the weight of the memory kernel reduces as the time interval becomes of the order of the inverse shear rate. The diminishing memory kernel weight can be incorporated in the schematic theory by replacing the vertex by a term like $4\lambda e^{-\dot{\gamma}(t-s)}$ as the exponential reduces the weight of the term when $(t-s)$ becomes of the order of $1/\gdot$. The exponential is a simple form, but a variety of other forms are possible \cite{fuchs03,brader09}. Thus the schematic equations for the two-point correlators become
\begin{align}
\label{shearRC}
\frac{\partial R(t,t_w)}{\partial t} =& -\mu(t) R(t,t_w) +\delta(t-t_w) 
+4\lambda  \int_{t_w}^t e^{-\dot{\gamma}(t-s)} R(t,s)C(t,s)R(s,t_w)\d s \nonumber\\
\frac{\partial C(t,t_w)}{\partial t} =& -\mu(t) C(t,t_w)+2T R(t_w,t) 
+2\lambda  \int_0^{t_w} e^{-\dot{\gamma}(t-s)} C^2(t,s)R(t_w,s)\d s 
+4\lambda  \int_0^t e^{-\dot{\gamma}(t-s)} R(t,s)C(t,s)C(s,t_w)\d s.
\end{align}

Similarly, the schematic equations for the susceptibilities are
\begin{align}
\frac{\partial \chi_R(t,t_w)}{\partial t} &= -\mu(t)\chi_R(t,t_w)+(1-\omega(t))R(t,t_w) 
+4\lambda\int_{t_w}^t e^{-\gdot (t-s)} \chi_R(t,s)C(t,s)R(s,t_w)\d s  \nonumber\\
&+4\lambda\int_{t_w}^t e^{-\gdot (t-s)} R(t,s)\chi_C(t,s)R(s,t_w)\d s 
+4\lambda\int_{t_w}^t e^{-\gdot (t-s)} R(t,s)C(t,s)\chi_R(s,t_w)\d s \nonumber\\
\frac{\partial \chi_C(t,t_w)}{\partial t} &= -\mu(t)\chi_C(t,t_w)+(1-\omega(t))C(t,t_w)  
+4\lambda\int_0^{t_w} e^{-\gdot (t-s)} C(t,s)\chi_C(t,s)R(t_w,s) \d s  \nonumber \\
&+2\lambda\int_0^{t_w} e^{-\gdot (t-s)} C^2(t,s)\chi_R(t_w,s)\d s 
+4\lambda\int_0^{t_w} e^{-\gdot (t-s)} \chi_C(t,s)R(t,s)C(t_w,s)\d s  \nonumber \\
&+4\lambda\int_0^{t_w} e^{-\gdot (t-s)} C(t,s)\chi_R(t,s)C(t_w,s)\d s 
+4\lambda\int_0^{t_w} e^{-\gdot (t-s)} C(t,s)R(t,s)\chi_C(t_w,s)\d s \nonumber \\
&+4\lambda\int_{t_w}^t e^{-\gdot (t-s)} \chi_C(t,s)R(t,s)C(s,t_w)\d s 
+4\lambda\int_{t_w}^t e^{-\gdot (t-s)} C(t,s)\chi_R(t,s)C(s,t_w)\d s \nonumber \\
&+4\lambda\int_{t_w}^t e^{-\gdot (t-s)} C(t,s)R(t,s)\chi_C(s,t_w)\d s \label{shearchiC},
\end{align}
\end{widetext}
with the functions $\mu(t)$ and $\omega(t)$, being the schematic forms of $K_1(t)$ and $\omega_k(t)$ respectively, are given as
\begin{align}
\mu(t) =& T + 6\lambda \int_0^t e^{-\gdot (t-s)} C^2(t,s)R(t,s)\d s \\
\omega(t)=& 12\lambda \int_0^t e^{-\gdot (t-s)} C(t,s)\chi_C(t,s)R(t,s)\d s \nonumber\\
&+6\lambda\int_0^t e^{-\gdot (t-s)} C^2(t,s)\chi_R(t,s)\d s.
\end{align}

As we have discussed earlier, it is expected that the system will reach a steady state under shear after a waiting time $t_w$ of the order of the inverse shear rate. Thus, imposing shear constraints the growth of the dynamic length scale in an aging system. This is exactly what we find from the numerical solution of the mode-coupling equations. In Fig. \ref{ch5_gdot_2} we show the solution for an imposed shear rate $\gdot=10^{-2}$ with value of the parameters $T=1.0$ and $\lambda=2.0$. Starting with the initial conditions corresponding to a high temperature liquid, we see that the three-point correlator $\chi_C(t,t_w)$ grows and then the growth saturates when $t_w\sim 10^2$. In the inset we show how the correlation function behaves. The two-point correlator also reaches a steady state after a certain waiting time that is of the order of the inverse shear rate. In Fig. \ref{gdot_3} we show the behaviour of $\chi_C(t,t_w)$ and $C(t,t_w)$ for the same value of temperature $T=1.0$ and $\lambda=2.0$ but for the shear rate $\gdot=10^{-3}$. In this case, the correlators reach their steady state when the waiting time becomes of the order of $10^3$. Since the relaxation time $t_r$ goes as $t_w^{0.8}$, these results imply for a sheared system under aging, the relaxation time will behave as $t_r\sim\gdot^{-0.8}$, a testable prediction.

Since activated hopping is ignored within mode-coupling theory, if the quench is below the transition point the aging system will reach equilibrium in the absence of shear when the waiting time becomes infinity.
The system will be trapped into one of the local minima and thus become non-ergodic. However, an arbitrarily small shear stirs the system and helps it come out of the local minimum, thus restoring ergodicity. Shear not only cuts off the relaxation time of the system, it also sets the dynamic length scale.

Shear doesn't affect the dynamics much when $t_w$ is much smaller than $\gdot^{-1}$. We have seen earlier that the aging dynamics can not be characterized by an evolving effective temperature. Therefore, even under shear, the evolution towards the steady state can not be characterized by an effective temperature, however, the steady state itself can be \cite{sarojunpub}. If we take the $t_w\to\infty$ limit of the eqs. (\ref{shearRC})-(\ref{shearchiC}), the resulting theory will describe the sheared steady state. It is possible to define an effective temperature $T_{eff}$ for this steady state. We find that $T_{eff}$ increases with $\gdot$ and a simple algebraic form gives $T_{eff}\sim \gdot^{1/4}$. The trend of $T_{eff}$ governed by this theory is quite encouraging and it is important to note that no equilibrium relation like the fluctuation-dissipation relation has been used in deriving this theory \cite{mythesis,sarojunpub}. However, the full generality of the theory, whether it can also be used to describe athermal systems, remains to be tested.

\section{Discussion and conclusion}
\label{coarsening_discussion}
In this work we have adapted and extended mode-coupling theory to describe the nonstationary states and show that the resulting theory captures the key features of emergence and growth kinetics of glassy domains starting from a liquid background. We have achieved this through a suitably defined susceptibility $\chi_C(t,t_w)$ analogous to the one in Ref. \cite{biroli06} for the equilibrium system and monitoring its growth following a quench to a low-temperature state. The peak-height of $\chi_C(t,t_w)$ has the interpretation of correlation volume and its growth with waiting time $t_w$ gives the domain growth of glassy order. We find that the glassy correlation volume grows as $t_w^{0.5}$; this is slower than the growth dynamics in conventional coarsening. We can extract a relaxation time $t_{peak}$ as the time where $\chi_C(t,t_w)$ has its peak and $t_{peak}$ grows as $t_w^{0.8}$. These theoretical findings are supported by simulation results of Ref. \cite{parisi99,kob97}. The broad features of the three-point correlator are in qualitative agreement with \cite{parsaeian08}. A very recent experimental study also sees the growth of glassy correlation volume and relaxation time with waiting time \cite{brun12}; a comparison of our theory to their findings in the appropriate range of temperatures and time scales would be very welcome.

Next we have obtained the equations of motion for the growth kinetics in a $p$-spin spherical spin-glass model. Even though the qualitative features of domain growth as function of waiting time is similar to those for the structural glasses, the correlation volume has a much slower growth ($\sim t_w^{0.13}$) in this case with $p=3$. We hope these results will encourage further studies of how the length scale of dynamic heterogeneity affects the domain growth of spin-glass order and the aging dynamics in spin-glasses in general. 

We have further extended our theory for an aging system under steady shear and found that an imposed shear-rate $\dot{\gamma}$ cuts off aging and coarsening at $t_w \sim 1/\dot{\gamma}$ in the glassy region and $t_w = \min(t_r,1/\dot{\gamma})$ in the fluid. As the relaxation time goes as $t_w^{0.8}$, $t_r$ or $t_{peak}$ should vary as $\dot{\gamma}^{-0.8}$ for a sheared system. 
Note that this result is not valid in the aging regime, but applies only when $t_w\sim \gdot^{-1}$. Since the system reaches steady state at that time, the result remains valid in the steady state. This is an interesting and testable prediction of the theory.
 
An important feature of the dynamics emerging from studying the three-point correlator is that the dynamics at various waiting times can not be described with an evolving effective temperature $T_{eff}$, contrary to what the study of the two-point correlator seems to suggest. Thus, for the case of an aging system under shear, the evolution of the system towards the steady state can not be described by an effective temperature, although the final steady state itself has an well-defined $T_{eff}$. In results to be presented separately \cite{sarojunpub}, we find that $(T_{eff}-T)\sim \gdot^{1/4}$ although this power law fitting form was somewhat {\it ad hoc}, the qualitative features agree well with numerical experiments \cite{haxton07,ono02}.

It would be interesting to see what the full $k$-dependent theory predicts. We had to schematicise the equations in order to obtain a numerically tractable form. Even in that case it takes quite a long time (e.g., 10-15 days for the data in Fig. \ref{largelambda}) to extract a significant waiting-time dependence for various functions. To do anything better than this, we need to find a better numerical algorithm, but it is not clear how to go about this important task.

MCT is applicable in a narrow regime at the onset of glassy transition. How seriously should we then take the results presented here? If the quench is below the mode-coupling transition, but above an ideal glass transition, say the Kauzmann temperature $T_K$ \cite{kauzmann48}, activated processes that are not present within MCT \cite{sarika08} should cut off the growth in an experiment or simulation. But typical simulations do not explore these asymptotically long time scales and thus, can be usefully compared to our MCT coarsening results \cite{saroj12}. A quench below $T_K$ will presumably give indefinite growth of a different length scale \cite{berthier_kob_2011,cammarota_biroli_2011,kurchan11} with a form not predicted by MCT.

\section{Acknowledgements}
We would like to thank G. Biroli, C. Cammarota, C. Dasgupta, B. Kim, N. Menon, K. Miyazaki, S. Nagel and S. Sastry for many useful discussions and enlightening comments. SKN would also like to thank TIFR Hyderabad, where part of this work was carried out, for hospitality. SKN was supported in part by the University Grants Commission and SR by a J.C. Bose Fellowship from the Department of Science and Technology, India.


\begin{thebibliography}{93}
\bibitem[{foo({\natexlab{a}})}]{footnote2}
\bibinfo{note}{To understand this point it helps to look at the correlation
  functions in equilibrium as shown in Fig. \ref{ss_chi3scaling} where scaling
  $\chi_C(t,t_w)$ by $\chi_C^P(t_w)$ and time by $t_{peak}$ led to data
  collapse for all the curves corresponding to various $\lambda$ or
  temperature. Thus, if each curve corresponds to a unique effective
  temperature, we should obtain data collapse.}
\bibitem[{foo({\natexlab{b}})}]{footnote1}
\bibinfo{note}{The numerical solution of the anisotropic mode-coupling
  equations are highly time consuming, but possible for the steady state
  theory\cite{kuni04,henrich09}. However, the situation becomes much worse in
  case of aging and solving the $k$-dependent aging equations even without
  shear becomes unrealistic.}
\end{thebibliography}

\end{document}